# Refractory Metal Nuggets – Formation of the First Condensates in the Solar Nebula

Kurt Liffman[1], Francesco C. Pignatale[2], Sarah Maddison[2], Geoffrey Brooks[2]
1. CSIRO/MSE, P.O. Box 56, Highett, Victoria 3190, AUSTRALIA
2. Center for Astrophysics and Supercomputing, Swinburne University of Technology, Hawthorn VIC 3122, AUSTRALIA

**ABSTRACT**

As gas flowed from the solar accretion disk or solar nebula onto the proto-Sun, magnetic pressure gradients in the solar magnetosphere and the inner solar nebula provided an environment where some of this infalling flow was diverted to produce a low pressure, high temperature, gaseous, "infall" atmosphere around the inner solar nebula. The pressure in this inner disk atmosphere was mainly dependant on the accretion flow rate onto the star. High flow rates implied relatively high pressures, which decreased over time as the accretion rate decreased.

In the first hundred thousand years after the formation of the solar nebula, accretional flow gas pressures were high enough to create submicron-sized Refractory Metal Nuggets (RMNs) – the precursors to Calcium Aluminum Inclusions (CAIs). Optimal temperatures and pressures for RMN formation may have occurred between 20,000 to 100,000 years after the formation of the solar nebula. It is possible that conditions were conducive to RMN/CAI formation over an eighty thousand year timescale. The "infall" atmosphere and the condensation of refractory particles within this atmosphere may be observable around the inner disks of other protostellar systems.

The interaction of forces from magnetic fields with the radiation pressure from the proto-Sun and the inner solar accretion disk potentially produced an optical-magnetic trap above and below the inner solar nebula, which provided a relatively stable environment in which the RMNs/proto-CAIs could form and grow. These RMN formation sites only existed during accretion events from the proto-solar disk onto the proto-Sun. As such, the formation and growth time of a particular RMN was dependent on the timescale of its nascent accretion event.

Observational evidence suggests that RMNs were the nucleation particles for CAIs. As a consequence, the observed bimodal distribution of $^{26}$Al in CAIs, where some CAIs have $^{26}$Al while others do not, is probably due to the injection $^{26}$Al during the short CAI formation period, where $^{26}$Al was not present when the first CAIs were formed.

# 1. Introduction

Some particles obtained from Comet Wild 2 (Brownlee et al., 2006) are similar to the Calcium Aluminum Inclusions (CAIs) (Zolensky et al., 2006) and chondrules (Nakamura et al., 2008) that are found in primitive meteorites. Some of the Stardust chondrules and CAIs were also enriched in solar $^{16}$O (ibid., Simon et al., 2008, McKeegan et al., 2011, see also Appendix A). This is evidence that these particles formed near the Sun and were then transported to the outer regions of the Solar System Protoplanetary Disk, which we will hence-forth call the Solar Nebula (SN).

Fundamental astrophysical theory indicates that half the gravitational potential energy present in SN will be dissipated in the boundary region between the star and the disk (Frank et al., 2002). This theoretical requirement, plus the observation that high speed, bipolar jet flows are ubiquitous in young stellar systems, prompted the idea that recycled, heated material from near the inner rim of the SN would, via the agency of a solar bipolar outflow, be found in meteorites and comets (Skinner, 1990, Liffman, 1992, Liffman and Brown, 1995, 1996, Shu et al.,1996). It was deduced that a major portion of the SN was reprocessed by jet flows and that such reprocessed material should make up a significant portion of primitive meteorites (Liffman and Brown, 1995).

Observational confirmation of this jet flow theory has been obtained from Spitzer Space Telescope (SST) data of the young solar-like star Ex Lupi, where crystalline forsterite grains are formed within 0.5 AU of this young star (Ábrahám et al., 2009) and the grains were subsequently transported at speeds of tens of kilometers per second away from the star, possibly by a disk-driven wind, (Juhász et al., 2012). SST observations of the young stellar system HOPS-68 strongly suggest that crystalline forsterite grains were formed close to the young star and then ejected into the circumstellar environment via an outflow, whereupon the forsterite grains probably rain back down onto the disk surrounding the young star (Poteet et al., 2011).

Theorists have suggested a number of other radial transport mechanisms besides jet flows, e.g., turbulent convection in the SN (Clarke and Pringle, 1988, Ciesla, 2009, Hughes and Armitage, 2010), radiation pressure (Vinković, 2009) and photophoretic force (Krauss et al., 2007). It is plausible that some or all of these mechanisms played some role in the outward radial transport of SN material, but, to date, the only available observational evidence supports the jet flow theory for the formation and transport of processed material in the SN.

The suggested possible heating mechanisms for such particles have been quite imaginative (Scott, 2007), but, in this study, we will be constrained by the observed formation of crystalline forsterite grains in the disk surrounding the young solar-like star EX Lupi (Ábrahám et al., 2009). These crystalline grains formed directly from condensation (Pignatale et al., 2011) or from the thermal annealing of amorphous dust grains in the surface layer of the inner disk due to heat from an accretion-induced outburst of stellar radiation. A number of researchers have also suggested that solar radiation was a potential heating mechanism for the formation of CAIs, most chondrules and other refractory particles ( Eisenhour et al., 1994, Shu et al., 1996, Liffman, 2009). In particular, Shu et al. (1996) suggested that the obscuration of the Sun due to the local geometry on the inner SN may have played a role in producing CAIs and chondrules (e.g., see Eqns (6) to (9) of Shu et al. (1996)). In this study, we wish to examine that hypothesis in more detail with the goal of understanding when and how the precursors to CAIs may have formed.

CAIs contain minerals that are high temperature condensates from a gas of solar composition (Marvin et al., 1970, Grossman, 1972, Simon et al., 2002), but they also show evidence for remelting and reprocessing (MacPherson, 2003). As such, CAI formation cannot always be modeled as a pure condensation sequence. The CAI formation process was more complicated than the initial model of CAI formation: a slowly cooling, hot nebula from which the CAIs condensed. A more recent suggested astrophysical setting for CAI formation was near the Sun (Shu et al., 2001), using the "X-wind" to form and transport the CAIs. The "X-wind" model for CAI formation has suffered a significant amount of criticism (e.g., Wood, 2004, Desch et al., 2010). None-the-less, the "near-Sun" CAI formation scenario remains attractive due to the high temperatures in that region and the source of solar $^{16}$O-enriched oxygen (see Appendix A for further details). Progress may be made in identifying the site for CAI formation if there are subcomponents of CAIs that are pure condensates, since the appropriate calculations would provide constraints for the formation of the condensates.

Equilibrium condensation calculations (Grossman, 1973) indicated that an alloy of the most refractory metals: W, Re, Os, Ir, Mo, Ru, Pt, and Rh, would probably be the first condensate from a solar gas (Palme and Wlotzka, 1976). Concentrated levels of these elements with relative abundances consistent with equilibrium condensation were found by Eisenhour and Buseck (1992) (see also Blander et al., 1980 and (Sylvester et al., 1990)) in five submicron-sized refractory metal nuggets (RMNs) enclosed in spinel grains. Berg et al. (2009) isolated 458 RMNs from the Murchison meteorite and chemically analyzed 88 RMNs. All the RMNs were submicron in size, seemingly pristine, and had compositions consistent with equilibrium condensation. For an assumed pressure of 10 Pa, Berg et al. deduced maximum cooling rates of around 1 K/year between the condensation temperature range of 1620K and 1450K, thereby suggesting formation timescales of order one hundred years.

In addition, CAIs contain some of the oldest materials to have formed in the SN and show abundant evidence that the SN was enriched in short-lived radionuclides such as $^{26}$Al (Scott, 2007). $^{26}$Al decays to $^{26}$Mg with a half-life of 720,000 years and its presence in CAIs follows a bimodal distribution with the initial values of $^{26}$Al/$^{27}$Al being ($^{26}$Al/$^{27}$Al)$_0$ ~ 5×10$^{-5}$ and ($^{26}$Al/$^{27}$Al)$_0$ ~ 0 (MacPherson et al., 1995, Simon et al., 2002, Makide et al., 2011). If one assumes that $^{26}$Al was homogenously distributed in the SN then a strict chronological interpretation would suggest an initial spike of CAI formation at or around the start of the SN with another formation spike of CAI some millions of years later (Podosek and Cassen, 1994). Alternatively, this bimodal distribution of $^{26}$Al may have been due to spatial heterogeneity, where some parts of the SN had $^{26}$Al and other parts did not (ibid). However, Sahijpal and Goswami (1998) argued for temporal heterogeneity, where the $^{26}$Al-free CAI formed prior to the presence of $^{26}$Al in the CAI formation region of the SN. In this model, CAIs formed over a relatively short length of time that happened to span the injection of $^{26}$Al into the SN. A CAI formation theory must provide an explanation of how this bimodal distribution of $^{26}$Al arose.

Given these constraints, the challenge is to find a region of the SN where such cooling rates and gas pressures could be achieved. To accomplish this goal, we first consider the physics and length scales of the inner SN (§2). The representative particle temperatures and gas pressures during an accretion event are then calculated for the gas at the base of the flowing columns of gas between the SN and the proto-Sun (§3), where it is shown that the ideal temperatures for the formation of RMNs, and hence also the CAIs, occurred soon after the SN was formed. In §4, the steady-state form of the ideal magnetohydrodynamic (MHD) equations are examined to compute the expected pressures for an atmosphere of gas surrounding the inner SN during an accretion event. The final section (§5) examines how radiation pressure and induced magnetic fields around the inner SN can trap sub-micron particles. Temperatures and temperature cooling rates of the RMNs are then dependent on the time scale over which accretion events occurred in the early SN.

## 2. The Inner Solar Accretion Disk

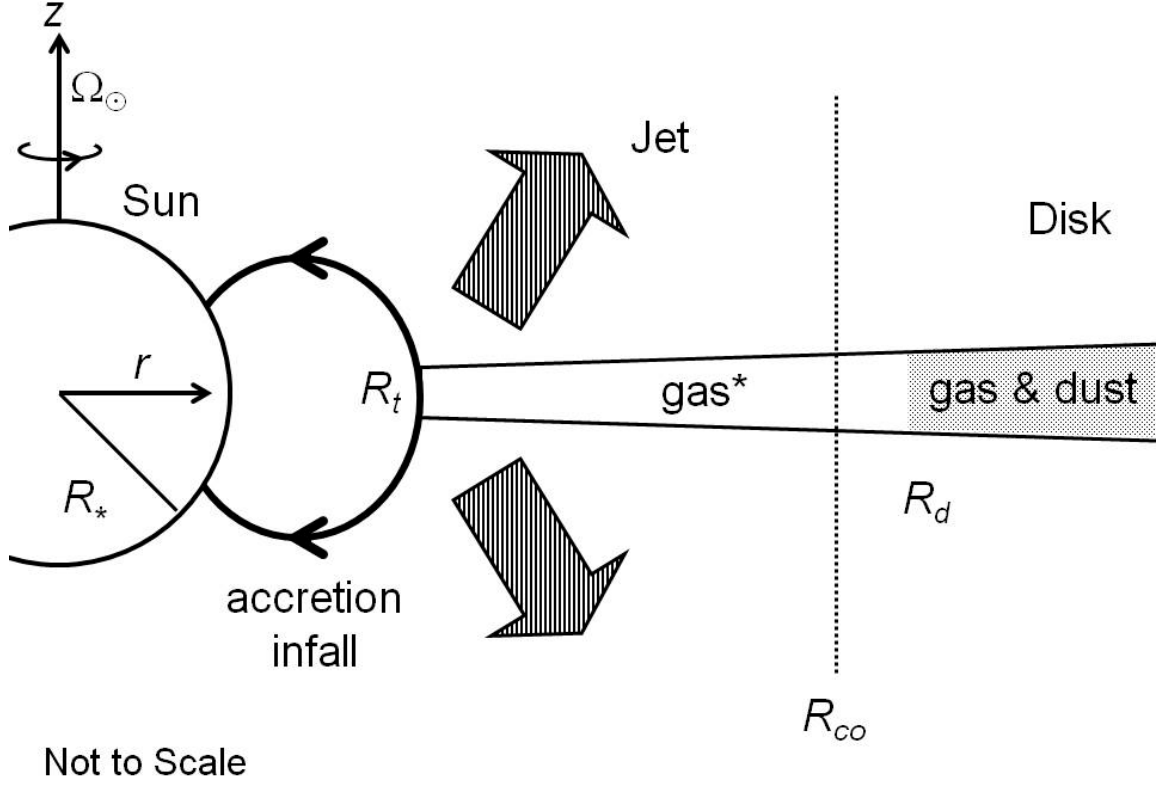

Figure 1: Length scales of the inner SN: $R_t$ - disk truncation radius, $R_*$ - radius of a star/Sun, $R_{co}$ - co-rotation radius, $R_d$ - dust sublimation radius and $\Omega_\odot$ - angular rotational frequency of the Sun, $r$ and $z$ are the standard cylindrical coordinates. Silicate dust particles are destroyed for $R < R_d$. The "gas*" refers to gas and refractory particles. The case of $R_t < R_{co}$ is shown, where mat1erial accretes along stellar magnetic field lines.

A schematic summary of the structure of the inner SN is given in Figure 1. Material from the inner disk accretes onto the star via stellar magnetic field lines (Königl, 1991). Here, $R_t$ is the distance of the inner edge of the disk from the center of a star, where the stellar magnetosphere truncates the accretion disk (Ghosh and Lamb, 1978)

$$R_t \approx \left( \frac{4\pi}{\mu_0} \frac{B_*^2 R_*^6}{\dot{M}_a \sqrt{GM_*}} \right)^{2/7} = 0.067 \left( \frac{(B_*(R_*)/0.1\,\text{T})^2 (R_*/2\text{R}_\odot)^6}{(\dot{M}_a/10^{-8}\,\text{M}_\odot\,\text{year}^{-1})(M_*/\text{M}_\odot)^{1/2}} \right)^{2/7} \text{AU}, \qquad (1)$$

where $\mu_0$ is the permeability of free space, $B_*$ is the stellar magnetic field strength, $R_*$ is the stellar radius, $\dot{M}_a$ is the accretion rate of material onto the star, $G$ the universal gravitational constant and $M_*$ the mass of the star. In Eqn (1), we have chosen a representative value of 0.1 T for the magnetic field strength on the surface of the proto-Sun. This value is consistent with observations, which suggest surface field strengths for proto-stars in the range 0.1 to 0.3 T (Güdel, 2007). We have taken the lower end of the magnitude range, because the surface magnetic fields are multipolar – often a mixture of dipole and octupole components (ibid., Gregory and Donati, 2011). The multipolar fields will rapidly decrease in magnitude with distance from the star relative to the dipole field. For example, octupole magnetic fields drop off as $(R_*/r)^5$, while dipole magnetic fields decrease as $(R_*/r)^3$. So the dipole field is likely to dominate at the inner edge of an accretion disk (Long et al., 2007, Güdel, 2007).

We assume that the mass accretion rate onto the star has the form:

$$\dot{M}_a(t) \approx \dot{M}_a(t_0)\left(\frac{t}{t_0}\right)^{-\eta}, \qquad (2)$$

with $t > 10^4$ year, $t_0 = 10^6$ year, $\eta = 1.5$ and $\dot{M}_a(t_0) \approx 4 \times 10^{-8}$ $M_\odot$ yr$^{-1}$. This very approximate formula is based on observed mass accretion rates (Hartmann et al., 1998). It should be noted that the above equation represents the average accretion rate. The actual accretion rate will very probably be subject to episodic accretion with significant variations around the mean value (Dunham et al., 2010).

The co-rotation radius, $R_{co}$, is the distance from the star where the angular rotation frequency of the star is equal to the Keplerian angular frequency, $\Omega_{Kep}$, of the accretion disk. If we assume Keplerian rotation for the disk then

$$\Omega_{Kep} = \left(\frac{GM_*}{R^3}\right)^{1/2} = 1.78 \times 10^{-5}\left(\frac{(M_*/M_\odot)}{(R/0.05\ \mathrm{AU})^3}\right)^{1/2}\ \mathrm{Hz}, \qquad (3)$$

and

$$R_{co} = \left(\frac{GM_*}{\Omega_*^2}\right)^{1/3} = 0.078\left(\left(\frac{M_*}{M_\odot}\right)\left(\frac{P_*}{8\ \mathrm{days}}\right)^2\right)^{1/3}\ \mathrm{AU}, \qquad (4)$$

where $\Omega_*$ is the angular rotational frequency of the star and $P_*$ is the rotational period of the star. In Eqn (4), we have normalized the rotation period to 8 days, which is, approximately, the median rotation period for ~ solar mass, pre-main sequence stars that are around $10^6$ years old (Herbst et al., 2007). In this study, we require values for the rotation rate of proto-Sun/protostars at ages of around $10^4$ and $10^5$ years. Unfortunately, such information is not available, so the observed pre-main sequence rotation periods are used as an indicative guide.

The relative difference in angular velocity between the disk and the co-rotating stellar magnetic field generates a current within the disk. In Figure 2, we show a section of the stellar/disk circuit. Here the current density generated within the disk, $j_D$, travels along the inner stellar magnetic field lines, $j_M$, and then returns to the disk via the outer stellar magnetic field lines in the corona above the disk, $j_C$. Only the current flow between $R_t$ and $R_{co}$ is shown. The full star-disk circuit is shown in Bardou and Heyvaerts (1996).

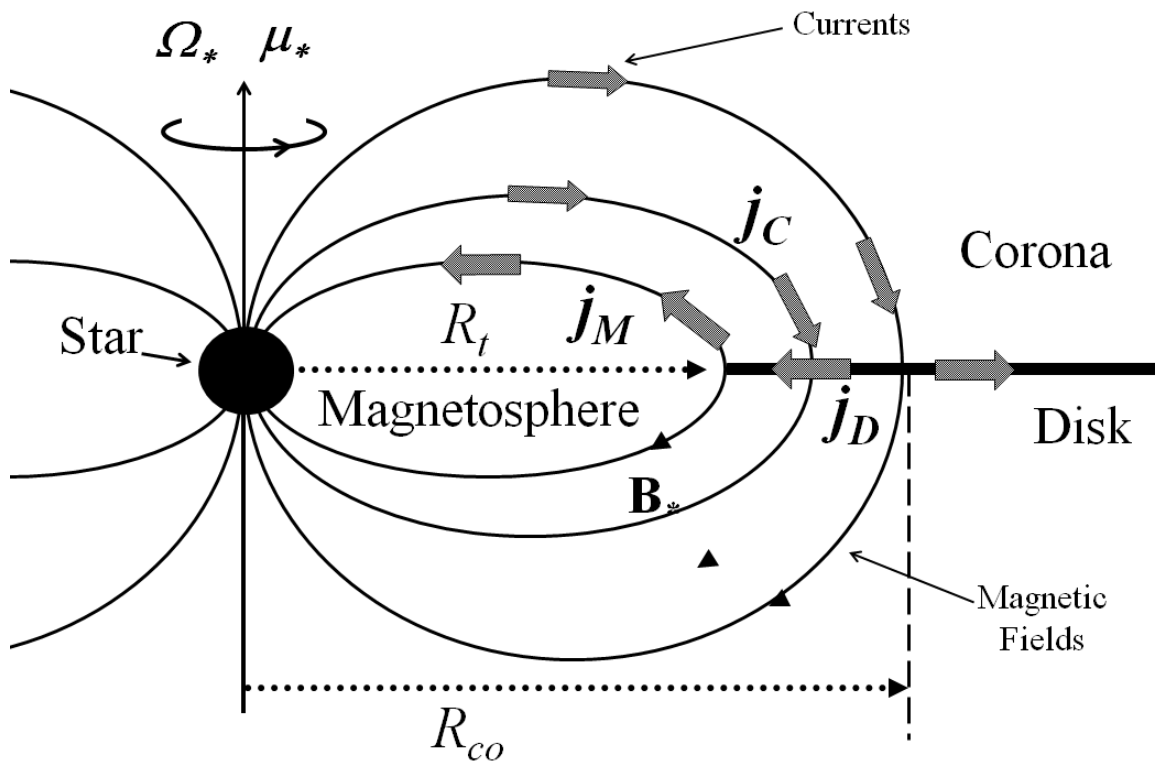

Figure 2: Current flows in the inner section of a star/disk circuit. $j_D$ - disk current, $j_M$ - field aligned stellar magnetosphere current, $j_C$-field aligned coronal currents. If the stellar magnetic field pointed in the opposite direction, to that shown in the figure, then the direction of the current flows would reverse. We show an approximately dipole field, but stellar fields are typically multipolar, where the dipole field tends to dominate near the disk.

Close to the star, the magnetic stellar field co-rotates with the star, so if $R_t < R_{co}$ then the angular velocity of the disk is greater than the star ($\Omega_{Kep}(R_t) > \Omega_*$) and a disk current, $j_D$, plus a todoidal magnetic field, $B_\phi$, is generated within the inner disk. The resulting $j_D \times B_\phi$ Lorentz force compresses the inner accretion disk (Campbell and Heptinstall, 1998, Liffman and Bardou, 1999) as is shown schematically in Figure 3. A similar disk profile is derived analytically and computationally in Campbell (2010), while Barnes and Macgregor (2003) used a magnetohydrodynamic model to derive the same electric current circuit.

The height of this magnetically compressed disk is approximately given by the magnetic disk height, $H_B$, (Liffman and Bardou, 1999)

$$H_B(r) = h(r)\sqrt{\ln(1 + \rho_c(r)/\rho_\infty)}, \qquad (5)$$

where $\rho_c(r)$ is the gas density in the central plane of the disc at a distance $r$ from the center of the central object, $\rho_\infty$ is proportional to the energy density of the magnetic field (*ibid.*) and $h(r)$ is the isothermal scale height of the disk:

$$h(r) = \sqrt{\frac{2r^3 k_B T_D}{GM_* \bar{m}}} = 0.0013\,\mathrm{AU}\sqrt{\frac{(r/0.05\,\mathrm{AU})^3 (T_D/1500\,\mathrm{K})}{(M_*/M_\odot)(\bar{m}/m_{H_2})}}, \qquad (6)$$

where $k_B$ is the Boltzmann's constant, $T_D$ the temperature of the inner disk, $r$ the cylindrical radial distance from the centre of the star and $\bar{m}$ the mean molecular mass of the gas. $H_B$ is not a scale height, but is the true surface of the magnetized disk with a sharp cutoff in the density for $|z| > H_B$ (*ibid.*). This implies that any atmosphere above the inner disk surface can be completely separated from the inner disk. For the cases of interest, $H_B$ is usually $< h(r)$.

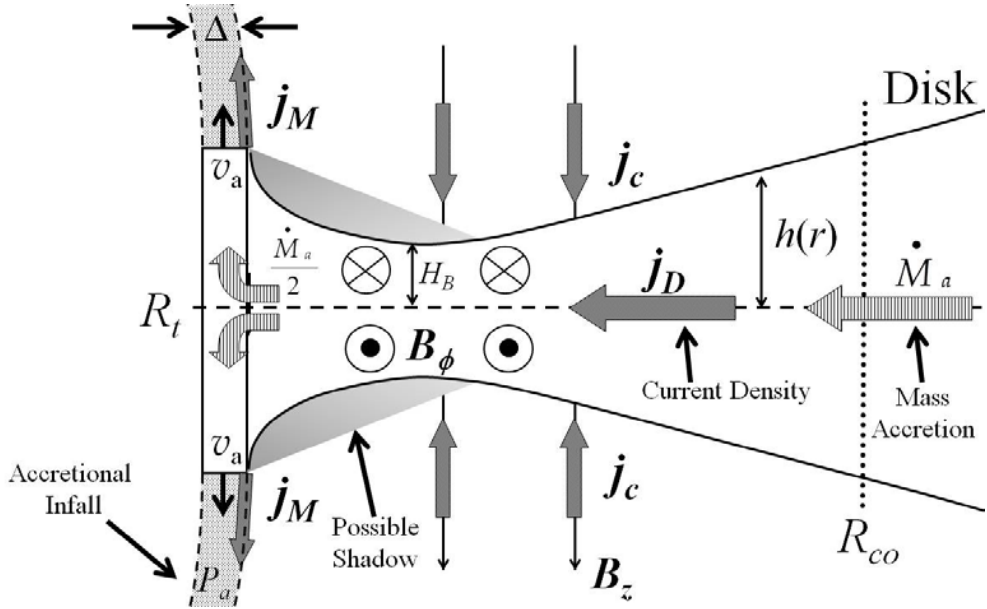

Figure 3: A schematic of the inner accretion disk region for $R_t < R_{co}$ that has a mass accretion rate $\dot{M}_a$. The relative motion between the inner disk and stellar magnetic field generates a radial current, $j_D$, and toroidal magnetic field, $B_\phi$, within the disk, which produce a Lorentz force that compresses the disk to a magnetic height, $H_B$. The circuit is completed by the currents $j_M$ and $j_c$ via the stellar magnetospheric field lines, $B_z$, between the star and the disk. At $R_t$, the compressive Lorentz force is removed and the disk expands and falls along initially vertical magnetic field lines to the stellar surface in a region of width $\Delta$ with a speed $v_a$. The gas in the infalling material has a pressure $P_a$. If the inner disk is optically thick then part of the compressed disk may be in shadow due to the infalling material. Alternatively, this system may be optically thin, but becomes optically thick due to the nucleation of a sufficiently high number of dust particles.

For $r < R_t$, the disk material decreases its angular velocity from Keplerian rotation to co-rotation with the star. As a consequence, the disk material begins to fall towards the star along the stellar magnetic field lines (Campbell 2010). This infalling material will be hot and in a mostly gaseous state. Indeed, it is the archetypical gaseous SN material of the first theories used to model the formation of the condensates in the SN.

## 3. Infalling Gas – A source of condensate material.

Near the inner disk, the infalling gas and dust will tend to flow along the stellar (approximately) dipole field lines with an initial speed, $v_a$, in the z direction in an (assumed) axisymmetric channel of initial width $\Delta$ (Figure 3). A number of authors have developed detailed and elegant flow models for the velocity and density of the infalling gas, e.g., Adams and Gregory (2012) and references therein. However, this is still a controversial field (Campbell, 2010, D'Angelo and Spruit, 2010) and we thought it prudent to compute some representative values for $v_a$ and $\Delta$. As such, we have adopted the standard boundary layer value for $\Delta$, where the stellar magnetosphere at $R_t$ replaces the surface of a compact object (e.g., Eqn (6.10) in Frank et al., 2002):

$$\Delta \approx \frac{h(R_t)^2}{R_t} = 2 \times 10^{-5} \text{AU} \frac{(h/0.001\text{AU})^2}{(R_t/0.05\text{AU})}. \tag{7}$$

From the conservation of mass.

$$\frac{\dot{M}_a}{2} = 2\pi R_t \Delta \rho_a v_a. \tag{8}$$

Combining Eqns (7) and (8) gives:

$$\rho_a = \frac{\dot{M}_a}{4\pi h(R_t)^2 v_a}. \tag{9}$$

The maximum initial infall velocity, in the z direction, may be computed from the conservation of energy and is given by

$$v_a = \sqrt{\frac{GM_*}{R_t} - R_t^2 \Omega_*^2}. \tag{10}$$

Eqn (10) assumes immediate, perfect energy transformation from excess angular velocity (i.e., material flowing from the Keplerian disk to the co-rotating magnetosphere for $R_t/R_{co} < 1$) to vertical flow. In reality, the infalling flow will start at subsonic speeds near the disk and arrive at supersonic speeds at the surface of the proto-Sun. None-the-less, for the mass flows and stellar parameters that are considered in this paper, we obtain a characteristic accretion infall gas number density, $n_a$, near the disk of

$$n_a = \frac{\rho_a}{\bar{m}_g} \approx 4.3 \times 10^{13}\,\text{cm}^{-3}\,\frac{\left(\dot{M}_a/10^{-6}\,\text{M}_\odot\,\text{yr}^{-1}\right)}{\left(h(R_t)/10^9\,\text{m}\right)^2 \left(v_a/30\,\text{km s}^{-1}\right)\left(\bar{m}_g/2.3\,\text{m}_\text{H}\right)}, \qquad (11)$$

where $\bar{m}_g$ is the average mass of the gas and $m_H$ is the mass of hydrogen. This representative value is of the same order of magnitude as that given in Adams and Gregory (2012).

The major constraints on RMN formation are the pressure and temperature of the SN gas from which the condensate forms. To compute the pressure, we need the temperature and density of the infalling gas. As a surrogate for the gas temperature, we use the effective particle temperature as the particle has to condense from the gas. As is discussed in Appendix B, the effective temperature of a particle is a function of the solar luminosity.

The luminosity of the proto-Sun has two components: the radiation from the proto-Sun, $L_*$, and the accretion luminosity, $L_a$. The accretion luminosity has the form

$$L_a = \frac{GM_* \dot{M}_a}{R_*}\left(1 - \frac{R_*}{2R_t}\right) = 15.4\,\text{L}_\odot\,\frac{(M_*/\text{M}_\odot)(\dot{M}_a/10^{-6}\,\text{M}_\odot\,\text{yr}^{-1})}{(R_*/2\text{R}_\odot)}\left(1 - \frac{R_*}{2R_t}\right). \qquad (12)$$

The values for the luminosity and radius of the early Sun may be obtained from a standard solar model (Figure 4). These values are still an approximation as the solar model assumes a constant solar mass. In this paper, our values of $L_a$ and $L_*$ implicitly contain an extinction coefficient:

$$L_{a,*} = L^o_{a,*} e^{-\tau}, \qquad (13)$$

with $L^o_{a,*}$ being the source values of $L_a$ and $L_*$, while $\tau$ is the optical depth of the gas between the Sun and the inner rim of the disk. We have set $e^{-\tau} = 1$ for most of this paper, but will change this criterion slightly in §5. Another exception is the disk itself, which we assume is optically thick. As a consequence, the compressed region behind the infalling disk in Figure 3 is shielded from the Sun up to an assumed distance from the midplane of $\sim h(R_t)$.

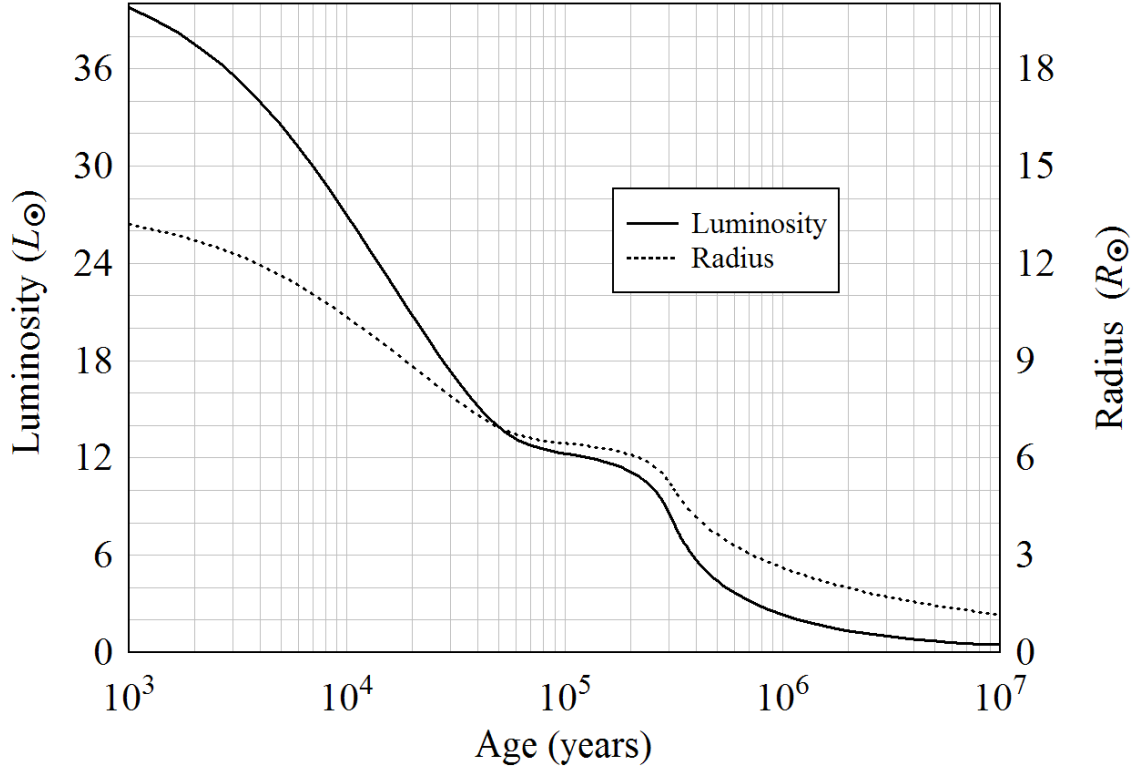

Figure 4: Luminosity and radius of a Sun-like star (X = 0.703, Y = 0.277, Z = 0.020) for the first 10 million years (Siess et al., 2000). The plateau region at around $10^5$ years is due to the onset and demise of deuterium burning.

For the special case of a particle that is not obscured by the disk and is not subject to diffuse radiation from the disk, the particle temperature $T_p$ is (see Appendix B)

$$T_{p-star-direct} = \left( \frac{(L_* + L_a)\varepsilon_a}{16\pi R^2 \sigma_B \varepsilon_e} \right)^{1/4} = 1250\,\text{K} \left( \frac{((L_* + L_a)/L_\odot)\varepsilon_a}{(R/0.05\,\text{AU})^2 \varepsilon_e} \right)^{1/4}. \tag{14}$$

In Eqn (14), the emission ($\varepsilon_e$) and absorption ($\varepsilon_a$) coefficients are usually dependent on the frequency of the radiation. Hence, the proportion of energy absorbed by the particle from the accretion radiation will be different relative to the energy absorbed from stellar radiation. In this study, we only wish to determine the approximate temperatures of the particle, so we assume that the emission and absorption coefficients are independent of the radiation frequency and set $\varepsilon_e = \varepsilon_a = 1$.

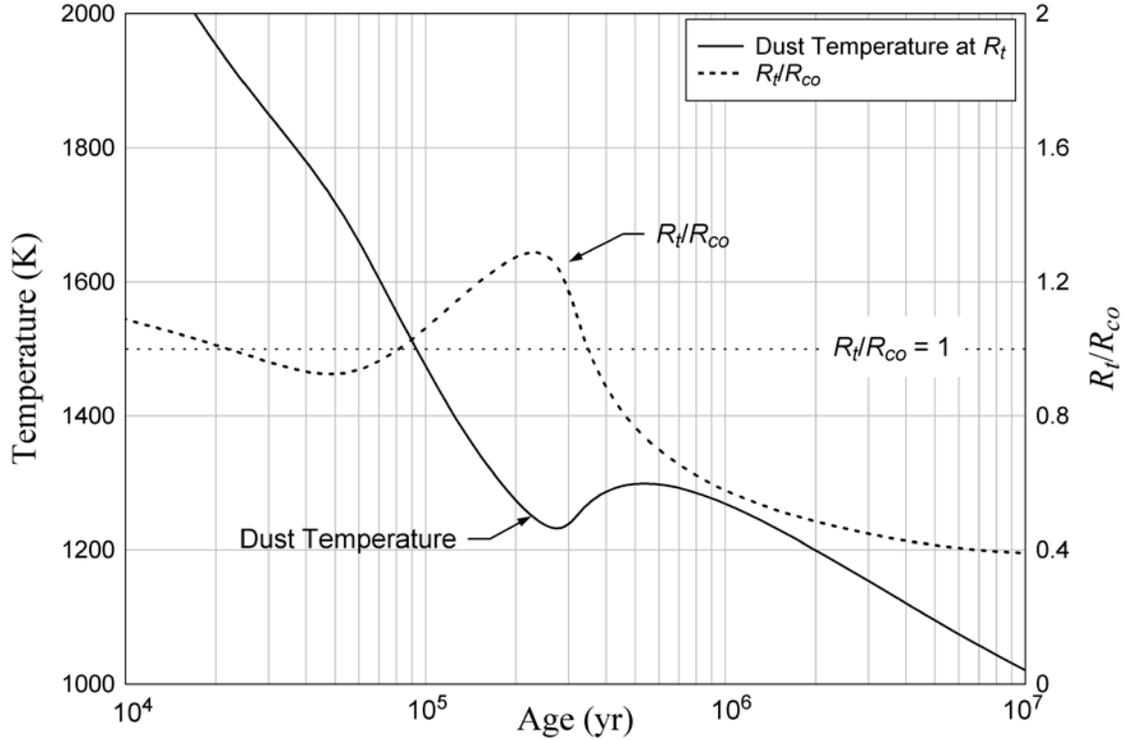

Figure 5: The solid line shows the approximate temperature of a particle (Eqn (14)) subject to direct radiation from the Sun at $R_t$, (Eqn (1)) the inner truncation radius of the accretion disk. $R_t/R_{co}$ is the ratio of the truncation radius to the co-rotation radius (Eqn (4)). Accretion can readily occur when $R_t/R_{co} <1$. Accretion is less likely to occur when $R_t/R_{co} >1$ (the deuterium burning phase). Temperatures are close to CAI and Ameboid Olivine Aggregate (AOA) formation temperatures when $R_t/R_{co} <1$ during the time range of 20,000 to 80,000 years after the beginning of the SN. Note that the second phase of $R_t/R_{co} <1$ accretion is consistent with the timescale between CAI and chondrule formation (Liffman, 2009).

A plot of Eqn (14) is given in Figure 5. As can be seen, between $10^4$ and $10^5$ years, the truncation radius is approximately less than or equal to the co-rotation radius and, as a consequence, accretion is more likely to occur. During this time, the particle temperatures are close to the RMN formation temperatures (Berg et al., 2009). After ~ $10^5$ years, deuterium burning starts in the proto-Sun and this keeps the radius of the proto-Sun relatively constant (Figure 4). At the same time, the accretion rate is still decreasing and so $R_t$ slowly increases. As a consequence, the truncation radius becomes greater than the co-rotation radius and accretion becomes less likely. After ~ $3\times10^5$ years, the deuterium has been converted to $^3$He and the contraction in the radius of the proto-Sun continues. As a consequence, the truncation radius becomes less than the co-rotation radius and accretion is assured. Liffman (2009) has suggested that this deuterium–induced variation in $R_t/R_{co}$ is the reason why there is a gap between the periods of CAI formation and chondrule formation.

For these calculations, we have taken the radius and luminosity values for the pre-main sequence Sun as a function of time as given by Siess *et al*. (2000). For our calculation of the inner radius of the disk as given by Eqn (1), we have used the mass accretion rates given by Eqn (2) and also calculated the increasing mass of the Sun over time as determined by the accretion rates. We have also assumed a solar rotational period of 18 days and a surface dipole magnetic field strength of 0.08 T (Herbst et al., 2007, Güdel, 2007).

It should be noted that the Siess et al. (2000) calculations, for the luminosity and radius of the pre-main sequence Sun, assume a constant solar mass with no mass accretion. Clearly, this was not the case for the proto-Sun, which would have built up in mass over time due to accretion from the SN onto the proto-Sun. The Siess et al. evolutionary tracks, quite probably, overestimate the radius of the proto-Sun during the protostar phase. As a consequence, we overestimate the temperature at the inner edge of the Solar Nebula. In an effort to overcome this, we use a larger $R_t$ to decrease the temperatures and because accretion can only occur if $R_t < R_{co}$, we also require a larger $R_{co}$, which in turn implies a rotation period that is longer than the median rotation period of ~8 days, but within the observed range of 2 to 20 days for pre-main sequence stars (Herbst et al., 2007). Of course, one possible solution for overestimating the rotation period is to obtain more realistic pre-main sequence tracks, which include mass accretion onto the star, but such data are not available at the time of writing.

Combining Eqns (9), (10) and (14), we can obtain an expression for the approximate pressure, $P_a$, for the infalling accreting gas in the near neighbourhood of the disk:

$$P_a = \frac{k_B \rho_a T_p}{\bar{m}_g} \approx \frac{k_B \dot{M}_a \left((L_* + L_a)\varepsilon_a\right)^{1/4}}{4\pi h (R_t)^2 \bar{m}_g \left(\frac{GM_*}{R_t} - R_t^2 \dot{\Omega}_*^2\right)^{1/2} \left(16\pi R_t^2 \sigma_B \varepsilon_e\right)^{1/4}}. \qquad (15)$$

In Figure 6, $P_a$ is shown as a function of time. There are gaps in the pressure record due to deuterium-burning phase when $R_t/R_{co} > 1$, which makes accretion less likely. The pressures obtained in the accreting gas around the 20,000 to 30,000 year mark are similar to the $10^{-4}$ atmospheres (10 Pa) suggested for CAI/RMN formation (Berg et al., 2009, Blander, 1979).

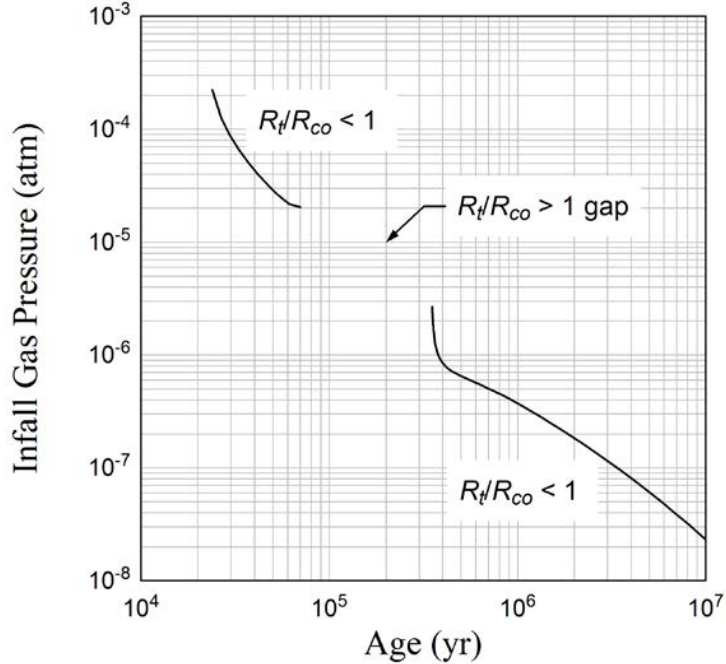

Figure 6: The pressure, $P_a$, of the infalling accreting gas as it leaves the inner radius of the accretion disk as a function of the age of the SN. The gaps in the pressure distribution are due to the deuterium-burning phase which gives $R_t/R_{co} > 1$ with a decreased probability of accretion. The pressures obtained between 20,000 and ~100,000 years are compatible with CAI and AOA formation.

These calculations suggest quite a short timescale: ~ 80,000 years, for when the pressures in the infalling gas were in the appropriate range for RMN/CAI formation. A timescale of ~80 kyr for the existence of an environment that was conducive to RMN/CAI formation is consistent with the recently deduced 20 kyr to 50 kyr results for timescales over which CAIs formed (Dauphas and Chaussidon, 2011).

A short timescale for CAI formation has a number of consequences: (i) the apparent bimodal distribution (i.e., $(^{26}Al/^{27}Al)_0 \sim 5\times10^{-5}$ and $(^{26}Al/^{27}Al)_0 \sim 0$) of live $^{26}Al$ in CAIs can only be explained by the $^{26}Al$-free CAI being formed prior to the presence of $^{26}Al$ in the CAI formation region of the SN (Sahijpal and Goswami, 1998) and (ii) the observed range in values of $(^{26}Al/^{27}Al)_0 \sim 5\times10^{-5}$ to $3\times10^{-5}$ suggests a heterogeneous spatial distribution of live $^{26}Al$ in the SN and/or subsequent high temperature reprocessing over a timescale of around 700,000 years, where the reprocessed CAIs re-equilibrated with the slowly decreasing, ambient $^{26}Al$ (MacPherson et al., 2012).

## 4. The Atmosphere of the Inner Accretion Disk

As discussed in Appendix C, the current density flows $j_M$ and $j_C$ (which becomes $j_z$ closer to the disk) shown in Figure 2, generate a toroidal magnetic field (*i.e.*, a field that lies in a plane that is parallel to the disk plane) above and below the inner accretion disk. The strength of the toroidal field decreases with distance from the midplane of the disk, if radial currents can deplete $j_M$ (*e.g.*, Barnes and Macgregor, 2003) and/or the solar magnetic lines change direction from the *z* direction near the disk to the *r* direction as one moves away from the disk midplane. This decreasing toroidal field produces a magnetic pressure gradient away from the disk, which may produce a buoyant atmosphere around the inner disk or, indeed, an outflow of gas away from the disk (Uchida and Shibata, 1985, Liffman, 2007).

The RMNs we observe in meteorites clearly did not form in the material that accreted onto the Sun. However, we suggest that some of this infalling gas was diverted from the accretional infall and created an atmosphere of high temperature gas around the inner rim of the solar accretion disk (Figure 7). To determine whether this is possible, we need to consider the possible magnetic and current density configuration around the inner accretion disk. We can then use the magnetohydrodynamic (MHD) equations to work out the expected gas density leakage from the infalling gas.

To compute the pressures of the gas in this hypothetical atmosphere we also require the gas temperatures. As RMNs are condensate particles, the appropriate temperatures of the gas are the effective particles temperatures (Appendix B). The resulting particle temperatures around the inner disk are shown in Figure 8. There is significant spatial variation in the temperatures, since the top (and bottom) of the inner SN is partially shaded by the accretion curtain of infalling material that is travelling along the solar magnetosphere (see Figure 1).

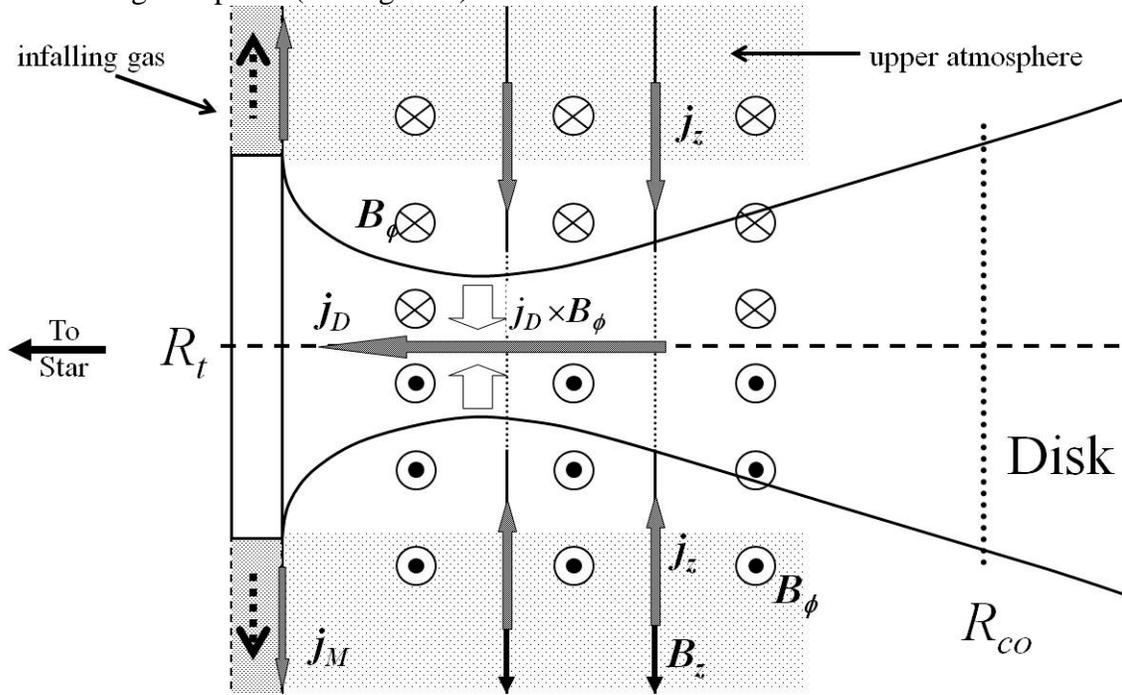

Figure 7: As gas is accreted onto the Sun, it moves along the solar magnetosphere and, initially, in a direction that is approximately perpendicular to the inner accretion disk. We show that some of this infalling gas is diverted and produces an atmosphere around the inner disk. It is from this outer atmosphere that RMN/CAI condense. Also shown are the current flows, magnetic fields and some of the resulting Lorentz forces. In particular, the radial current $j_D$ which interacts with the toroidal fields to compress the inner disk.

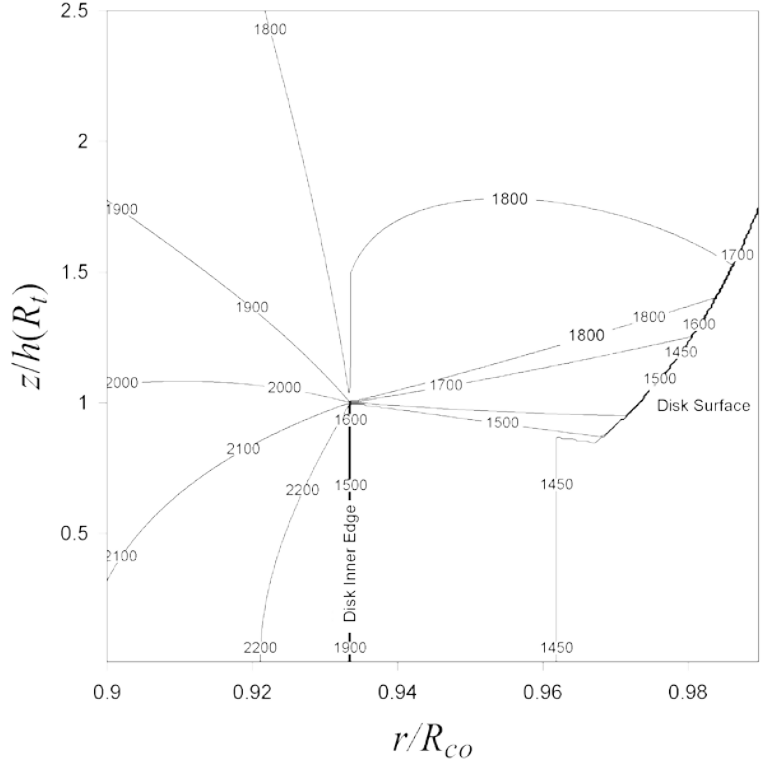

Figure 8: Contour plot of the effective particle temperature (K) around the inner accretion disk (Appendix B) at an age of 70,000 years from the start of the Solar System. The $z$ coordinate has been normalized to the isothermal scale height at the truncation radius. The $r$ coordinate has been normalized to the co-rotation radius ($R_{co}$ = 0.121 AU). The inner truncation radius ($R_t$ = 0.113 AU) is located at ~ $r/R_{co}$ = 0.934. The luminosity of the proto-Sun is $L_*$ ~ 12 $L_\odot$ and the radius is ~ 6.5 $R_\odot$. The accretion luminosity is $L_a$ ~ 6.7 $L_\odot$.

To determine the gas density above and below the inner disk, we use the steady state MHD momentum equation

$$\rho(\mathbf{v}\cdot\nabla)\mathbf{v} = -\nabla p - \rho\nabla\phi_g + \mathbf{j}\times\mathbf{B} , \qquad (16)$$

where $\mathbf{B}$ is the magnetic field vector ($\mathbf{B} = (B_r, B_\phi, B_z)$), $\mathbf{j}$ the current density is given by Ampere's Law:

$$\mathbf{j} = \frac{\nabla\times\mathbf{B}}{\mu_0} , \qquad (17)$$

with $\mu_0$ the permeability of free space and $\phi_g$ is the gravitational potential:

$$\phi_g = -\frac{GM_*}{\sqrt{r^2 + z^2}} . \qquad (18)$$

We note that using the steady state form of the MHD equations removes the possibility of magnetic instabilities destroying the computed density profiles. As such these computations are undertaken in the spirit of exploration as a proposed new site for RMN and CAI formation. The actual validity of the solutions has to be determined by more detailed, time dependent calculations.

The "$r$" component of the MHD momentum is

$$\rho\left(v_r\frac{\partial v_r}{\partial r}+v_z\frac{\partial v_r}{\partial z}-\frac{v_\phi^2}{r}\right)=-\frac{\partial p}{\partial r}-\frac{GM_*r\rho}{\left(r^2+z^2\right)^{3/2}}+j_zB_\phi-j_\phi B_z, \quad (19)$$

If we assume that $z \ll r$, and $v_r \approx 0$ then Eqns (19), (17) and the ideal gas law (Eqn (15)) imply

$$\frac{\partial\rho}{\partial r}+\frac{\rho\overline{m}}{k_BTr}\left(v_K^2-v_\phi^2\right)+\frac{\overline{m}}{k_BT\mu_0}\left(\frac{B_\phi^2}{r}+B_\phi\frac{\partial B_\phi}{\partial r}+B_z\frac{\partial B_z}{\partial r}\right)\approx 0. \quad (20)$$

with

$$v_K \approx v_\phi, \quad (21)$$

and, at the midplane of the disk (which is assumed to be aligned with the equator of the proto-Sun)

$$B_z(r)\approx B_z(R_*)\left(\frac{R_*}{r}\right)^3, \quad (22)$$

where $v_K$ is the Keplerian velocity (Eqn (D6)), $R_*$ is the radius of the star, $B_z(R_*)$ is the value of the $z$ component of the stellar field at the stellar surface. Eqn (21) assumes that most of the relevant physics occurs in the magnetic shear layer between the disk and the magnetosphere (Barnes and Macgregor, 2003).

The toroidal field can be described by

$$B_\phi(r)\approx\frac{\mu_0 I(r)}{2\pi r}. \quad (23)$$

$I(r)$ is the total current that is flowing to and from the disk in the vertical direction (Figure 3) for either the top or bottom halves of the disk:

$$I(r)=f_{Mz}I_M-I_z(r). \quad (24)$$

$I_M$ is the total current that is generated by the interaction between the solar magnetosphere and the top (or bottom) half of the disk (Eqn (D5) and Appendix D). The factor $f_{Mz}$ is the proportion of $I_M$ that travels in the $z$ direction. The simulations of Barnes and Macgregor (2003) suggest that $f_{Mz} \lesssim 1$, where some of the current generated in the disk may move radially across the magnetospheric fields lines and return to the proto-Sun. In this study, we have set $f_{Mz} = 1$ to examine the behavior of the system in a classical current configuration.

The formula for the current density $j_z$ as shown in Figure 7 is given by the formula (Liffman, 2007):

$$j_z(r) = -\sigma_D(r)h(r)\Omega_K(r)B_{*z}(r)\left(\left(\frac{r}{R_{co}}\right)^{3/2} - \frac{5}{2}\right).  \tag{25}$$

$I_z(r)$ is the summation of $j_z$ for $r > R_t$:

$$\begin{aligned} I_z(r) &= \int_{R_t}^{r} j_z(r) 2\pi r \, dr \\ &= \frac{2\pi\sqrt{GM_*}B_z(R_*)R_*^3}{\mu_0 \alpha c_s R_t^{5/2}}\left(\left(1-\left(\frac{R_t}{r}\right)^{5/2}\right) + \left(\frac{R_t}{R_{co}}\right)^{3/2}\left(\left(\frac{R_t}{r}\right)-1\right)\right). \end{aligned} \tag{26}$$

To compute the gas pressure above the inner disk, we also need to consider the $z$ component of the MHD equation (Eqn (16)):

$$\rho\left(v_r\frac{\partial v_z}{\partial r} + v_z\frac{\partial v_z}{\partial z}\right) = -\frac{\partial p}{\partial z} - \frac{GM_*z\rho}{\left(r^2+z^2\right)^{3/2}} - \frac{1}{2\mu_0}\frac{\partial B_\phi^2}{\partial z}. \tag{27}$$

We again consider the static and isothermal approximation, thereby obtaining:

$$\frac{\partial\rho}{\partial z} + \frac{\bar{m}GM_*z\rho}{k_BT\left(r^2+z^2\right)^{3/2}} + \frac{\bar{m}}{2k_BT\mu_0}\frac{\partial B_\phi^2}{\partial z} = 0. \tag{28}$$

We would expect that $\frac{\partial B_\phi^2}{\partial z} < 0$, so Eqn (28) indicates that the toroidal magnetic field will support the gas vertically above the disk, while the Solar gravity will tend to drain the gas towards the disk midplane.

We now have enough information to numerically solve for the gas density. To start the simulation, we set our time to 70,000 years, which sets, via Eqn (2), a total mass accretion rate of $2.2\times10^{-6}$ $M_\odot$ yr$^{-1}$. The mass of the proto-Sun is then found to be 0.74 $M_\odot$, the stellar surface magnetic strength is 0.1T and the rotational period of the proto-Sun is 18 days. From Figure 4, the luminosity of the proto-Sun is ~ 12 $L_\odot$ and the radius is ~ 6.5 $R_\odot$. The resulting gas pressure in the accreting gas along the solar magnetosphere (Eqn (15)) is $P_a = 3.8\times10^{-5}$ atm or 3.85 Pa, while $R_t = 0.113$ AU (Eqn (1)) and $R_{co} = 0.121$ AU (Eqn (4)). The accretion luminosity is ~ 6.7 $L_\odot$ (Eqn (12)) and $\alpha = 0.1$ (Eqn (D3) and Lasota, 2001).

Using the radial density obtained from Eqn (20) for the height $h(R_t)$, we then used Eqn (28) to provides us with an estimate of the gas density over the inner disk. Eqn (28) was solved by assuming that the toroidal magnetic field decreased linearly with $z$ over a length scale of the proto-Sun radius (6.5 $R_\odot$). This length scale is suggested by the dipole form of the solar magnetosphere and by the simulations of Barnes and Macgregor (2003), where the vertical component of the magnetospheric current, $I_M$, decreases on this length scale.

Using the ideal gas formula with the computed gas densities from the above equations plus the temperatures given in Figure 8, while assuming a gas with solar abundances (i.e., $\bar{m} \approx 2.3 m_H$) gives a pressure range of up to 70 Pa (~$7 \times 10^{-4}$ atm) above the inner disk (Figure 9) .

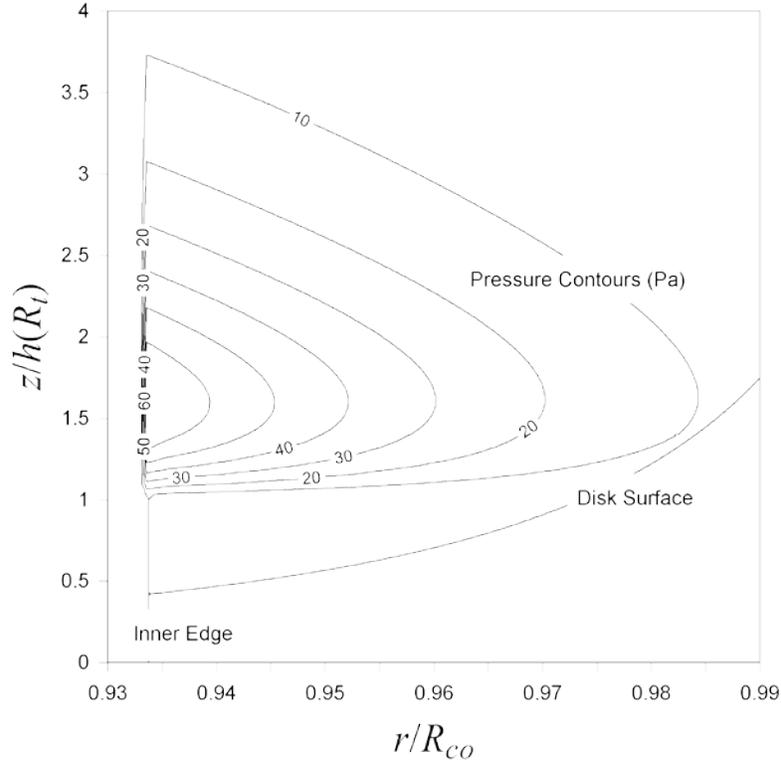

Figure 9: Estimated pressure (Pa) above the inner accretion disk due to radial leakage of gas accreting onto the star. The *z* coordinate is normalized to the isothermal scale height at the truncation radius, while the *r* coordinate is normalized to the co-rotation radius.

## 5. A Magnetic/Radiation Pressure Trap

RMNs possibly formed over a timescale of order one hundred years (Appendix E). RMNs are, in general, submicron-sized particles. This prompts the question of how such particles could remain for one hundred years in the active inner edge of an accretion disk of a very young stellar system, where one has infall onto the star and also outflow activities?

One class of possible constraint forces is due to radiation: photophoresis and radiation pressure. Photophoretic forces arise due to radiation producing an uneven temperature distribution over the surface of a particle (Krauss and Wurm, 2005). The warmer regions of the particle surface eject gas molecules at higher average speeds relative to the cooler regions and the difference in momentum transfer produces a force that is, usually, in a direction that is pointing away from the radiation source. Radiation pressure is due to the momentum carried by photons. In the case of submicron particles, such as RMN, the photophoretic force will, in most cases, be very small relative to the force due to radiation pressure (*ibid.*).

The magnitude of the radiation pressure force, $F_{rad}$, on a particle is given by:

$$F_{rad} \approx \int_0^\infty \Im(r,z) \frac{\pi a^2}{c} Q(\lambda, n, a) d\lambda ,\qquad (29)$$

where $\Im(r,z)$ is the radiative flux (Wm$^{-2}$), $a$ the radius of the particle, $\lambda$ the wavelength of radiation, $c$ the speed of light, $n$ the refractive index of the particle and $Q$ the radiation pressure efficiency factor (Wilck and Mann, 1996). The rate of energy per unit time, $q$, that is incident on the particle is simply

$$q(r,z,a) = \Im(r,z) \pi a^2 . \qquad (30)$$

The particle will receive radiative energy from the Sun, the inner solar nebula and the diffuse radiation from the gas above the inner solar nebula. The values of $q$ from the disk and the Sun are given in Appendix B. To compute the radiative flux from the gas above the disk requires computation of the density and temperature of this gas which is beyond the scope of this study and will not be given here. The direction of the radiation pressure forces from the disk and proto-Sun as a function of position is derived in Appendix B. Unfortunately, we do not have values for $Q$ or $n$ for RMN. So the accurate determination of $F_{rad}$ will have to await a separate study. Mie theory (Gindilis et al., 1969) suggests that $Q$ is dependent on the parameter

$$s = \frac{2\pi a}{\lambda} . \qquad (31)$$

From Wien's Law, the wavelength for peak radiation production is inversely proportional to the temperature. The temperature of the inner disk rim is approximately 1500K (Figure 8), while the temperature of the photosphere, as given by

$$T_* \approx \left( \frac{L_* + L_a}{4\pi R_*^2 \sigma_B} \right)^{1/4} , \qquad (32)$$

is approximately 4,700 K (assuming $t = 70,000$ years, which implies $L_a \approx 6.7 L_\odot$, $L_* \approx 12 L_*$ and $R_* \approx 6.5 R_\odot$). So, the peak wavelength (from Wien's Displacement Law) for the proto-Solar radiation is approximately 0.6 micron, while the peak wavelength for the inner disk is around 2 micron. $Q$ tends to peak strongly when the wavelength is around the size of the particle and since the observed RMNs have submicron sizes, we will make the plausible assumption that $Q_{Sun} \sim 1$ and set $0 \leq Q_{disk} \lesssim 0.1$.

Based on the geometry of our model protosolar inner accretion disk and the radiation pressure equations derived in Appendix B, the expected radiation pressures on the RMN are shown in Figure 10. Not surprisingly, the radiation pressure is mainly in the $r$ direction in a direction away from the proto-Sun. The accretion curtain, produced by infalling material travelling along the solar magnetosphere from the solar accretion disk to the proto-Sun, is assumed to produce a shadow region which limits the radiation pressure force near the surface of the accretion disk.

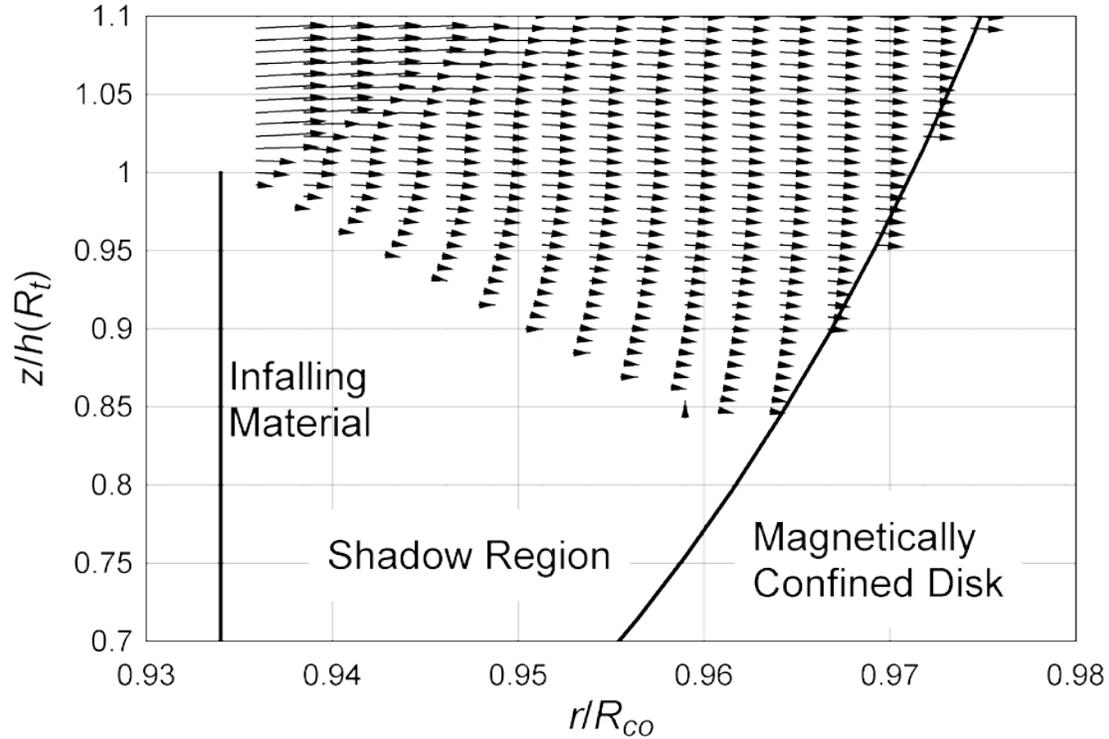

Figure 10: Radiation pressure vectors for radiation from the proto-Sun. It is assumed that the infalling material is optically thick up to the isothermal scale height of the disk at the truncation radius. This optically thick "curtain" produces a shadow region, where a portion of the proto-Sun is partially or completely obscured.

In addition to radiation pressure, there are also forces due to the interaction between the RMN's and the magnetic fields around the inner solar accretion disk. Such Lorentz forces arise because RMN's will gain a charge from the gas and the photo-ejection of electrons due to solar radiation (Appendix F). For the gas and radiation configurations given in Figure 9 and Figure 10, we obtain a charge distribution for RMNs (*i.e.*, the number of charges on each RMN) with a radius of 0.3 micron as shown in Figure 11. In this simulation, we are assuming a pure hydrogen gas. All of the RMN particles above and within the disk will have a negative charge, since the electron has a smaller mass and higher thermal speed relative to the protons.

In this example, we are assuming that the lowest gas pressure is around $10^{-3}$ Pa. The actual value of the charge is independent of pressure and is only dependent on the particle radius and temperature (Eqn (F8)), at least until the minimum gas pressure decreases to around $10^{-6}$ Pa at which point a positive charge is produced on the RMNs that are exposed to solar radiation.

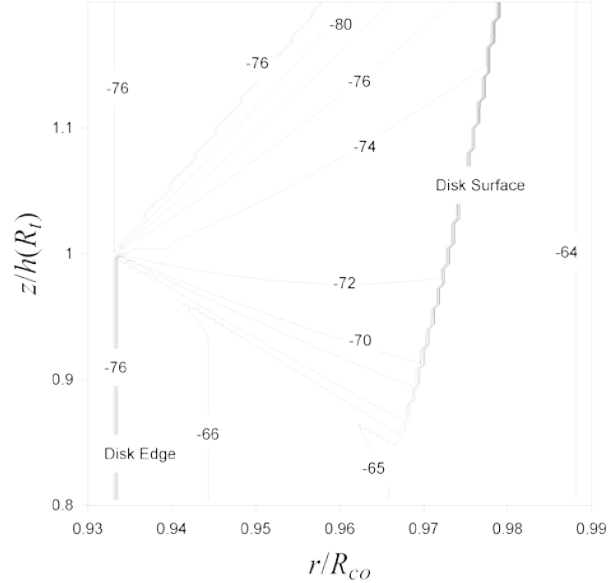

Figure 11: RMN charge distribution, $Z_d$, (Appendix F), *i.e.*, the number of charges on an RMN as a function of position above and in the inner solar accretion disk as derived for the gas and radiation distribution obtained from Figure 9 and Figure 10. The RMNs are negatively charged due to the higher thermal speed of electrons relative to protons.

When charged particles are subject to a force and are immersed in an electromagnetic field they tend to undergo particle drift (Northrup, 1963). A major source of drift motion, in this case, is due to external forces such as radiation pressure and gravity. Such force driven drift is described approximately by the equation:

$$\mathbf{v_F} \approx \frac{\mathbf{F_{ext}} \times \mathbf{B}}{qB^2} \ . \tag{33}$$

For this example, the radiation force is approximately a thousand times greater in magnitude relative to the *z* component of the proto-Sun's gravitational force (we ignore the *r* component of solar gravitational force as it is balanced by the centrifugal force in the rest frame of the particle). The radiation pressure vectors point away from the proto-Sun (Figure 10), while the toroidal magnetic fields above the disk point into the page (Figure 7). The resulting velocity drift vectors are shown in Figure 12 (the largest of which are of order $10^3$ ms$^{-1}$) do not account for gas drag. The radiation forces have been computed assuming $Q_{Sun} = 1$ and $Q_{disk} = 0.1$.

The drift velocities point towards the midplane until they reach the shadow region (due to an assumed optically thick accretion curtain) or the disk surface. In the shadow region the radiation force goes to zero and the particles are then held in place by the toroidal field. The same motion towards the midplane also occurs on the "underside" of the disk.

The *z* component of the gravitation field of the proto-Sun does not move the RMN closer to the disk because, by Eqn (33), it produces a drift in the azimuthal direction (*i.e.*, parallel to the toroidal field). If the RMNs reach a high enough number density then they can form a stable lattice-like structure (Fortov et al., 2004). Indeed, at high number densities the RMN region can become optically thick and the shadow region produced by an accretion curtain is no longer required. The mean free path for a photon interacting with 0.3 micron radius RMN is

$$l = \frac{1}{n\pi r_p^2} = 3,530 \,\text{km} \left(n/1\text{cm}^{-3}\right)^{-1} \left(r_p/0.3\,\text{micron}\right)^{-2}, \qquad (34)$$

So, given the distances within the inner accretion disk, we require the number density, $n$, of RMN to satisfy the condition $n > 10^4$ m$^{-3}$ to make the RMN region optically thick.

The charged RMN is held in place by the toriodal field. As such, the deduced cooling time scale of 1 K/year (Berg et al., 2009) for these particles may arise as the accretion event slowly decreases and the inner edge of the disk contracts back towards the co-rotation radius.

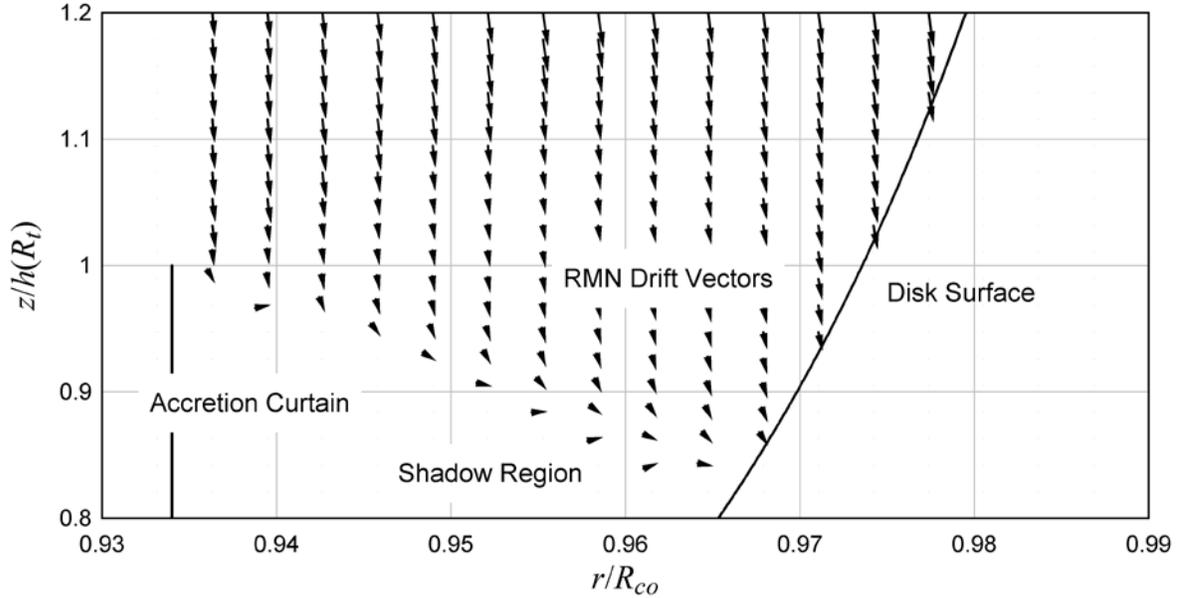

Figure 12: The RMN drift velocity vectors due solar radiation and a toroidal magnetic field. The RMNs drift towards the disk midplane until the solar radiation is blocked or until they reach the disk surface. The RMN would become locked into the shadow region until the inner disk moves back to the corotation radius and the accretion event is finished.

If the magnetic polarity of the proto-Sun were to be in the opposite direction to that shown in Figure 2 then the proto-RMN would be ejected from the system. For our RMN formation scenario to work we require the proto-Sun to undergo magnetic field reversals in a somewhat similar fashion to the modern Sun. During one cycle the RMN could form, while if an accretion event occurred during a period of opposite magnetic polarity then the RMN would not form.

# 6. Discussion

Recent observational results have demonstrated that solar radiation transforms amorphous dust into high temperature crystalline dust in a region close to a protostar (Ábrahám et al., 2009) and that these dust grains subsequently move radially away from their formation region at speeds of tens of kilometers per second, where the transport mechanism may be a jet flow (Juhász et al., 2012). Poteet et al. (2011) also use jet flows as a transport mechanism for crystalline grains that appear to form close to a protostar and rain back onto the protostellar disk.

These recent observations are consistent with the predictions of a number of authors who have suggested that most of the high temperature materials that we observe in meteorites and comets were formed in the inner SN and subsequently transported to other regions of the SN via the agency of a solar bipolar jet flow. In particular, these authors predicted that CAIs were also formed and transported via this process (Skinner, 1990, Liffman and Brown, 1995, 1996, Shu et al., 1996, 2001, Itoh and Yurimoto, 2003).

In this study, we have attempted to add some specific details to this general idea. In particular, we have introduced the new concept that RMNs and proto-CAIs were condensates from accretional infalling gas that was diverted from accretion columns by magnetic pressure gradients to form a high temperature, low pressure, "infall" atmosphere enveloping the inner SN. The RMNs and proto-CAIs were held in place in the infall atmosphere via an optical-magnetic trap that initially moved the particles towards the midplane of the inner SN, but then held them in place via the magnetic forces, once solar radiation was blocked by surrounding optically thick material. The general idea of forming proto-CAIs above the inner SN is consistent with the observed enrichment of Sun-like $^{16}$O in such condensates, where they formed outside the SN in the extended magnetosphere of the Sun from a mixture of $^{16}$O-rich SN gas and vaporized $^{16}$O-poor SN solids. A schematic view of this concept is given in Figure 13 and Figure A2.

We have shown that there was a finite time period of approximately 100,000 years where the pressures and temperatures in the infalling gas were high enough to form RMNs with condensation timescales that are physically reasonable. This places potential constraints on the mechanism that injected $^{26}$Al and other short-lived radionuclides into the SN. Due to this relatively short formation timescale, it is likely that $^{26}$Al was injected during the formation period of RMNs and CAIs.

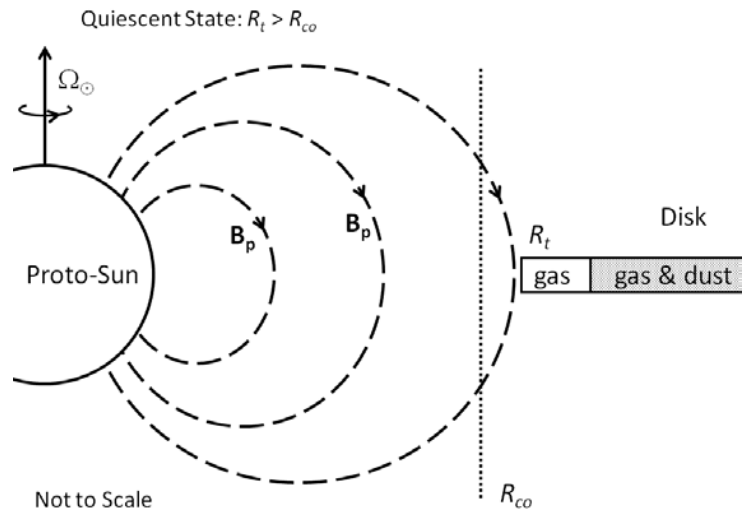

(a)

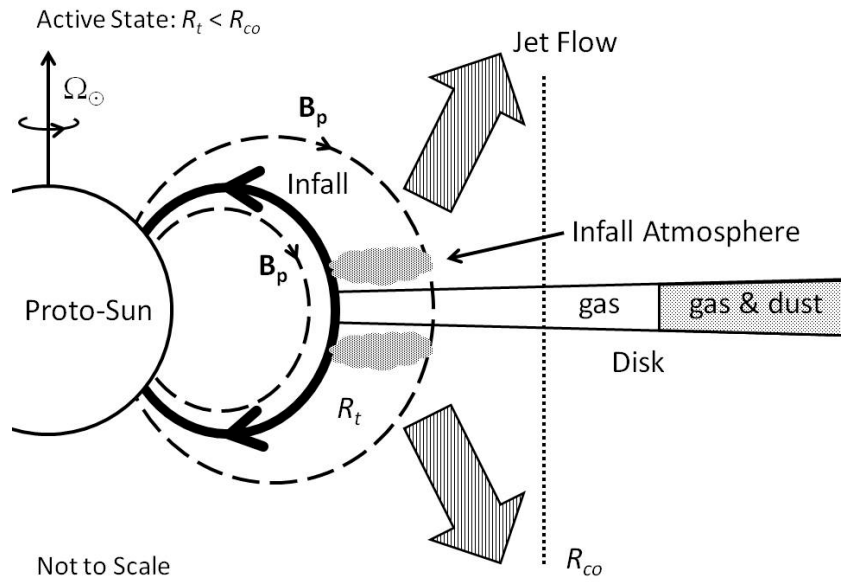

(b)

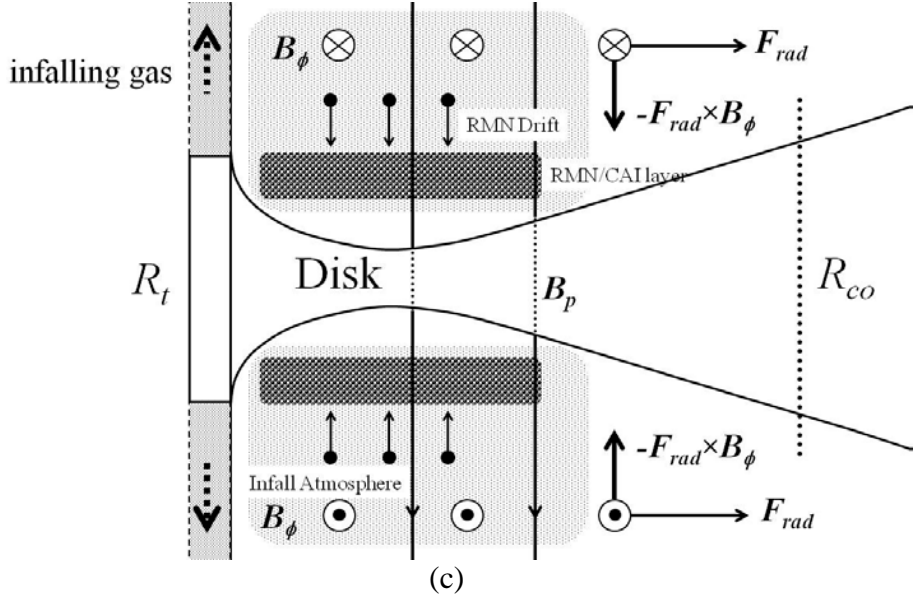

(c)

Figure 13: Stages in RMN/proto-CAI formation (a) Quiescent (Propeller) State: The truncation radius, $R_t$, is greater than the co-rotation distance, $R_{co}$, as produced by the poloidal field , $B_p$, of the solar magnetosphere. As a consequence, material is less likely to accrete onto the proto-Sun. (b) Active State: Now $R_t < R_{co}$ and material accretes onto the proto-Sun. A high temperature, low pressure "Infall Atmosphere" is formed from the infalling gas above and below the inner Solar Nebula. (c) RMNs/proto-CAIs condense from the Infall Atmosphere and gain a negative charge, whereupon they move towards the midplane of the Solar Nebula due to the radiation pressure force, $F_{rad}$, interacting with an induced toroidal field, $B_\phi$. This velocity drift stops when the radiation is blocked by the infalling gas or the self-shielding of the RMNs. The charged RMNs will be held in place by the poloidal field and will slowly cool as the accretion rate slowly decreases and the inner disk moves back towards the co-rotation radius.

## 7. Conclusion

We have examined the hypothesis that the formation of refractory metal nuggets (RMNs) and Calcium Aluminum Inclusions (CAIs) is intimately linked to the mass accretion from the Solar Nebula (SN) onto the proto-Sun. Our calculations indicate that within the first 100,000 years after the SN was formed, the pressures near or at the base of the accretion column that extended from the SN to the proto-Sun were comparable with the canonical pressures for the formation of RMN and CAI of around $10^{-4}$ atm.

Magnetic pressure from the solar dipole moved gas away from the accretion column to form a ~ 10 Pa, high temperature, $^{16}$O-rich atmosphere above the inner SN. Such an atmosphere provides the long sought after site from which RMNs and CAIs may have condensed. The RMNs became negatively charged in this "Infall Atmosphere" and were eventually held in place via the toroidal field induced by the interaction of the solar magnetosphere with the inner SN. As the accretion event slowly decreased, the RMNs were slowly cooled as the truncation radius of the inner SN slowly moved away from the proto-Sun

Such a short formation timescale for RMNs and CAIs can only imply that the, now extinct, $^{26}$Al found in some, but not all, CAIs was injected into the SN during this brief formation period. Some CAIs formed prior to and some after $^{26}$Al injection, thereby explaining the dichotomy of observed $^{26}$Al in CAIs.

As a consequence of this study, we predict the existence of an accretional "infall atmosphere" around the inner rim of solar-like protostellar disks. We also predict that refractory materials are forming in these infall atmospheres. Due to current resolution limitations, these predictions cannot be observationally tested. This situation may change with advances in observing technologies.

In summary, our model can explain:
   (i) The canonical pressures for the formation of RMNs and CAIs of around $10^{-3}$ to $10^{-4}$ atmospheres – this was due to the initial high rate of mass accretion from the Solar Nebula to the proto-Sun, which in turn produced the infall atmosphere around the inner solar nebula from which the RMN and CAI condensed;
   (ii) The short formation period for RMNs and CAIs of less than 80,000 years – again, this is linked to the mass accretion rate. In principle, it might be possible to produce some RMNs and CAIs whenever there was accretion from the SN to the proto-Sun. However, if the accretion rate was too low then the pressure in the infall atmosphere was also low. Such low pressures increase the time required to form RMNs and CAIs. If their formation timescale was much longer than the timescale of the mass accretion event then any RMNs and subsequent CAIs would be small in size and number relative to the RMNs and CAIs produced at the start of the Solar System;
   (iii) The $^{16}$O-enriched oxygen in CAIs – arose due to the solar-like gas in the infall atmosphere from which RMNs and CAIs condensed; and
   (iv) The dichotomy of $^{26}$Al in CAIs where some CAIs were exposed to $^{26}$Al, while others were not – was a function of the short RMN and CAI formation period and the injection of $^{26}$Al into the SN from a source that is still to be identified.

Despite these potentially hopeful aspects of the theory, there are still many uncertainties in the model and these will only be overcome with high resolution observations and numerical simulations of solar-like protostars, where the latter also has a detailed model for the interaction between a proto-stellar magnetosphere and the surrounding accretion disk.

# Acknowledgements

We acknowledge the generous support of the Australian Telescope National Facility (ATNF) Astrophysics Group, CSIRO Astronomy and Space Science Division and the Swinburne Centre for Astrophysics & Supercomputing. We wish to thank the anonymous referees for their constructive criticisms and express our appreciation to S. B. Simon and L. Siess for their comments and suggestions.

**Appendix A**

**Oxygen in the Solar Nebula**

Oxygen has three stable isotopes: $^{16}O$, $^{17}O$ and $^{18}O$. In the Solar System, the relative abundances of these stable oxygen isotopes have been measured to high precision in a large number of different materials (Ireland, 2012). The O isotopic ratios are usually expressed as permil deviations $\delta^{17}O$ and $\delta^{18}O$ from standard mean ocean water (SMOW):

$$\delta^i O = 10^3 \left( \frac{\left(^i O/^{16}O\right)}{\left(^i O/^{16}O\right)_{SMOW}} - 1 \right), \tag{A1}$$

where i = 17 or 18.

On a plot of $\delta^{17}O$ against $\delta^{18}O$ the Sun is enriched in $^{16}O$ by ~ 6% relative to terrestrial oxygen isotopes (McKeegan et al., 2011). All samples from the Earth, the Moon, Mars, and asteroid parent bodies lie very close, within 0.1%, to the terrestrial fractionation line (TFL): a line passing through the zero point (i.e., terrestrial) with a slope of approximately 0.52. This line arises due to equilibrium and kinetic processes that depend on the difference in mass between the oxygen isotopes (Young et al., 2002).

In contrast, the major components of primitive chondrites - such as CAIs, chondrules, AOAs, and fine-grained matrices - plot along a slope one line that connects the Earth and the Sun compositions (Clayton et al., 1973; Young and Russell, 1998). In most cases, CAIs and AOAs are more $^{16}O$-rich than fine-grained matrices and chondrules (Yurimoto et al., 2008). The unity slope of this line suggests it is a mixing line between a $^{16}O$-rich reservoir and a $^{16}O$-poor reservoir. The formation mechanism for these two reservoirs remains controversial (e.g., Ireland 2012) and will not be discussed here.

A metric for determining the amount of $^{16}O$ enrichment is given by

$$\Delta^{17}O = \delta^{17}O - 0.52\delta^{18}O, \tag{A2}$$

this is the distance from the TFL, and it provides a measure of $^{16}O$ enrichment unaffected by mass-dependent fractionation.

The Sun is significantly enriched in $^{16}O$ relative to the Earth with $\delta^{18}O_\odot \approx -58‰$, $\delta^{17}O_\odot \approx -59‰$, and $\Delta^{17}O_\odot = -28.4 \pm 1.8‰$ (McKeegan et al., 2011). Most of the meteoritical evidence suggests that the majority of solid material in the SN had $\Delta^{17}O_{solid} \sim 0$. Chemical equilibrium calculations show that for temperatures less then 1400K approximately 25% of the oxygen was in the solid phase and the remainder was in the gas phase (Figure A1). As a consequence, the SN gas had to be, on average, enriched in $^{16}O$. This is because the solar gas (with a value of $\Delta^{17}O_\odot \approx -28.4‰$) is a mixture of nebula solids and nebula gas, so by mass balance the SN gas was, on average, enriched in $^{16}O$ with a value $\Delta^{17}O_{gas} \approx \frac{4}{3}\Delta^{17}O_\odot \approx -37‰$. Such a value is comparable to the $\Delta^{17}O \approx -36‰$ from the Acfer 214 chondrule a006 (Kobayashi et al., 2003), although somewhat more $^{16}O$ enriched relative to the $\Delta^{17}O \approx -34‰$ found in some Isheyevo CAIs (Gounelle et al., 2009).

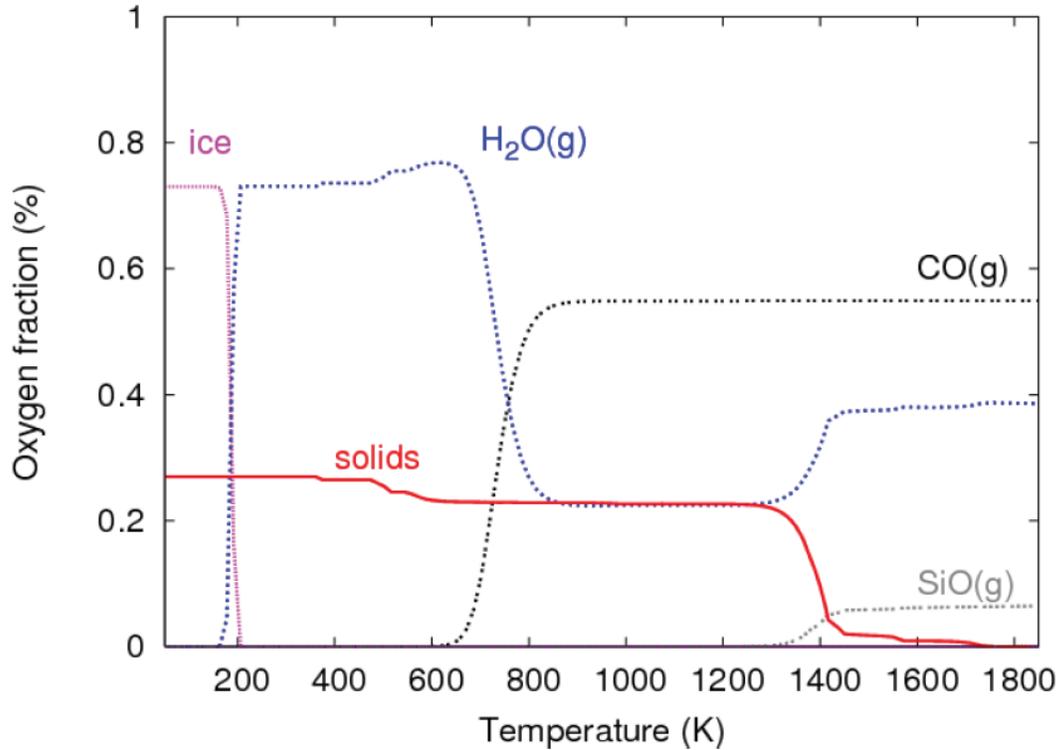

Figure A1: Chemical equilibrium calculations for oxygen compounds in the SN as a function of temperature with an assumed pressure of $10^{-4}$ atm (Pignatale et al., 2011). For temperatures less than 1400K, ~25% of the oxygen is in the solids with the remainder in the gas.

The $^{16}O$ spatial mapping that is suggested in this paper is shown in Figure A2. For convenience, we show the case where dust destruction radius, $R_d$, is less than the co-rotation radius, $R_{co}$ - although the same results apply if $R_d > R_{co}$. For $r > R_d$, the solids are $^{16}O$-poor and the gas is $^{16}O$-rich. For $r < R_d$, the solids start evaporating and release $^{16}O$-poor gas. The high temperatures at the truncation radius, $R_t$, ensure that most, if not all, the dust has evaporated and the gas has reached a solar value in the relative isotopic abundances of oxygen. It is this "solar value" gas that falls onto the proto-Sun and produces the infall atmosphere around the inner SN. As argued elsewhere in this study, RMN and CAI condense from this infall atmosphere and therefore it has the solar value of oxygen isotopes.

It will be noted that in Figure A2 we have given $\Delta^{17}O_\odot \approx -24‰$ at the time of RMN/CAI formation, while the current observed value is $\Delta^{17}O_\odot \approx -28.4‰$. The former value is consistent with the measured $\Delta^{17}O$ from CAIs and AOAs (Krot et al., 2010). It is also consistent with the idea that the SN became enriched with $^{16}O$ over time due to the solar bipolar jet flow ejecting $^{16}O$-enriched solids that landed back into the SN (Liffman 2009).

As a consequence, at the time of CAI formation, the SN solids and gas possibly had a slightly higher ($\sim 4‰$) $\Delta^{17}O$ value relative to decreasing values during the chondrule formation period, which occurred one to four million years after the CAI formation period (Villeneuve et al., 2009).

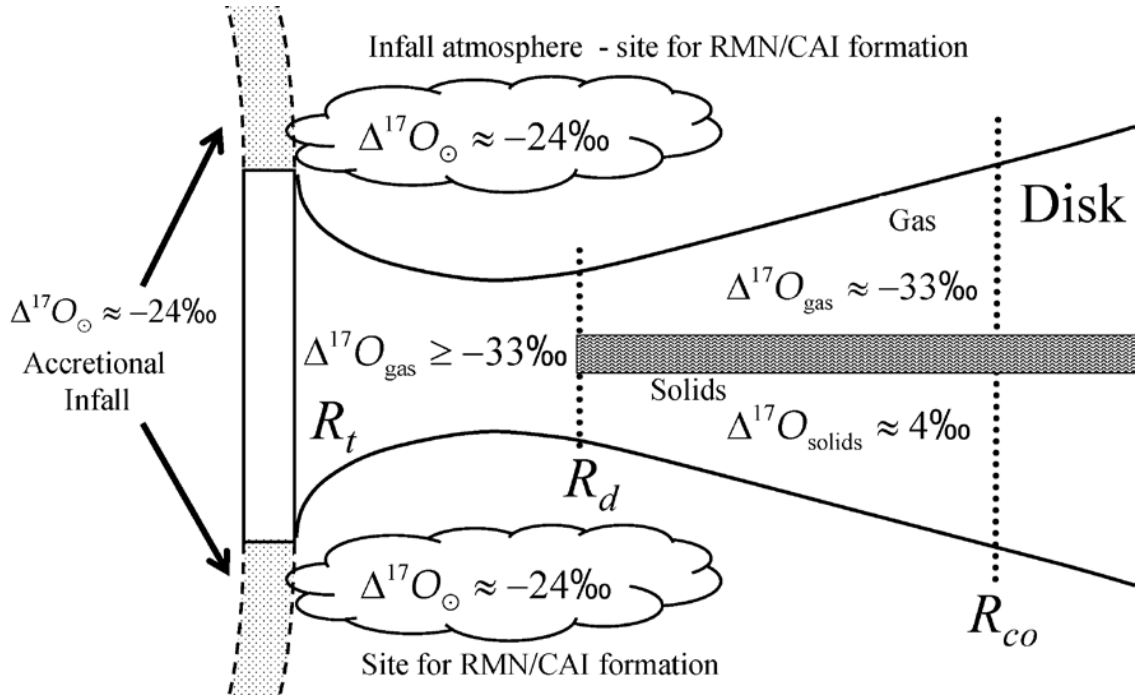

Figure A2: Suggested spatial map of $^{16}O$ enrichment in the inner SN at the time of RMN/CAI formation. $R_d$ is the dust destruction radius. For $r > R_d$, the dust is $^{16}O$-poor and the gas is $^{16}O$-rich. For $r < R_d$, the destruction of the dust decreases the relative amounts of $^{16}O$. The high temperatures at the truncation radius, $R_t$, ensure that all the dust has evaporated and that the gas has reached a solar value of $\Delta^{17}O_\odot \approx -24$. This gas subsequently forms the infall atmosphere around the inner disk from which RMNs and CAI form.

## Appendix B
## Particle Effective Temperature

### Stellar Radiation

In Figure B1, a spherical particle of radius, $a$, is located at a position ($r$, $z$) in the near neighborhood of the inner rim of an accretion disk, which has a height from the midplane denoted by $h$. The inner rim of the disk is located at a distance $R_t$, from the center of a star of radius $R_*$. The angles, $\phi$, $\theta_h$, $\chi$ shown in Figure B1, are defined by the equations:

$$\phi = \tan^{-1}\left(\frac{z}{r}\right), \tag{B1}$$

$$\theta_h = \tan^{-1}\left(\frac{h}{R_t}\right), \tag{B2}$$

$$\text{and } \chi = \tan^{-1}\left(\frac{z-h}{r-R_t}\right). \tag{B3}$$

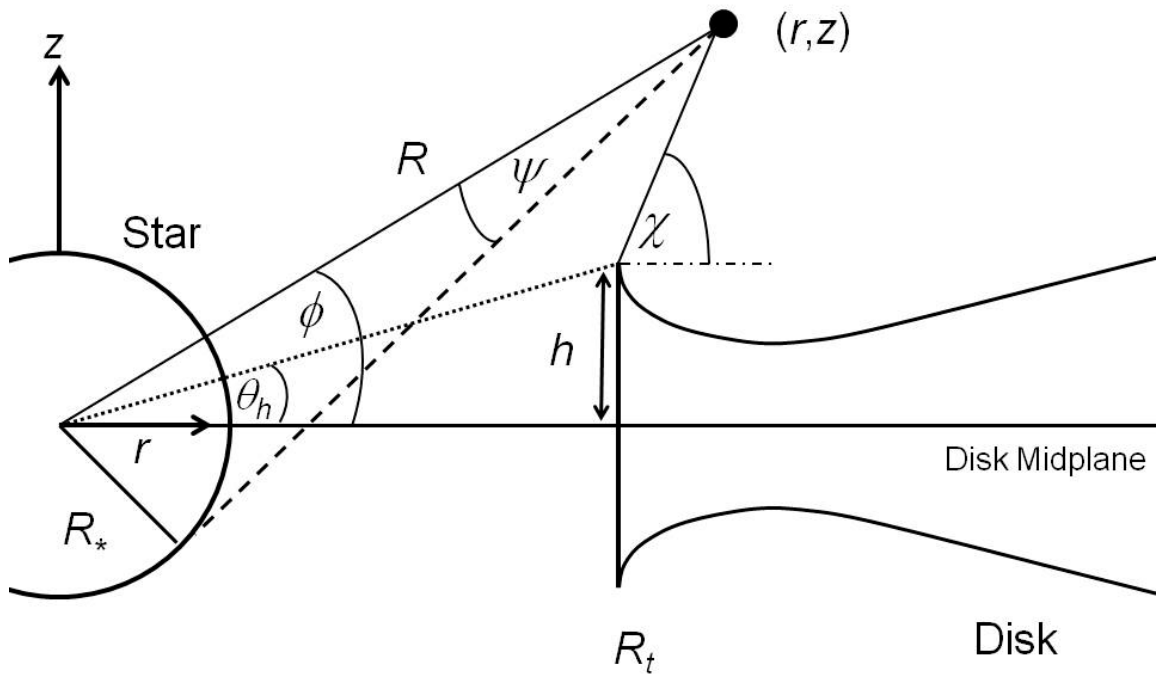

Not to Scale

Figure B1: A spherical particle is located at the point ($r$, $z$), which is in the near neighborhood of the inner rim of an accretion disk, If the particle is located close to the surface of the disk then its view of the star may be occluded and its effective temperature

will be lower relative to a particle that has a clear view of the entire star. The angle $\psi$ is the angle for the line of sight to the star. The other angles are defined in the text.

**Case 1: $\phi \geq \theta_h$**

For this case, the minimum of the star that the particle can "see" is half the stellar disk. As such, the amount of energy per unit time that will come from the star, $q_{star}$, is

$$q_{star} \approx \frac{(L_* + L_a)a^2 \varepsilon_a}{8R^2}\left(1 + \left(\frac{R}{R_*}\right)^2 \sin^2 \psi_l\right), \tag{B4}$$

where

$$0 \leq \psi_l = \chi - \phi \leq \psi_m = \sin^{-1}\left(\frac{R_*}{R}\right), \tag{B5}$$

with $\varepsilon_a$ the radiation absorption coefficient and

$$R = \sqrt{r^2 + z^2}. \tag{B6}$$

**Case 2: $\phi < \theta_h$**

In this case, less than half of the stellar disk is visible to the observer and

$$q_{star} \approx \frac{(L_* + L_a)a^2 \varepsilon_a}{8R^2}\left(1 - \left(\frac{R}{R_*}\right)^2 \sin^2 \psi_l\right), \tag{B7}$$

with

$$0 \leq \psi_l = \phi - \chi \leq \psi_m = \sin^{-1}\left(\frac{R_*}{R}\right). \tag{B8}$$

Disk Radiation
**Case 1: $z > h$**

When the particle is located above the disk, it will obtain direct stellar radiation from the star and diffuse radiation from the disk. For the case of a perfectly flat disk, it is possible to deduce from geometrical arguments that the radiation from the top of the disk is given by:

$$q_{disk-top} \approx 2\left(\frac{\pi}{2} + \tan^{-1}\left(\frac{r - R_t}{z - h}\right)\right)a^2 \varepsilon_a \sigma_B T_d(r)^4, \tag{B9}$$

where $\sigma_B$ is the Stefan-Boltzmann constant and $T_d(r)$ is the temperature of an optically thick, flat disk subject to stellar radiation (Friedjung 1985, Adams and Shu, 1986, Hartmann, 1998) and differential friction in the accretion disk (Frank et al., 2002 ):

$$T_d(r) \approx \left( \frac{(L_* + L_a)}{4\pi^2 \sigma_B R_*^2} \left( \sin^{-1}\left(\frac{R_*}{r}\right) - \frac{R_*}{r}\sqrt{1 - \left(\frac{R_*}{r}\right)^2} \right) \right.$$
$$\left. + \frac{3GM_* \dot{M}_a}{8\pi r^3 \sigma_B}\left(1 - \left(\frac{R_*}{r}\right)^{1/2}\right) \right)^{1/4} . \quad (B10)$$

For $r < R_t$, we set $T_d(r) = T_d(R_t)$.

In our case, we have a magnetically compressed disk between $R_t$ and $R_{co}$, so for particles with $R_t < r < R_{co}$,

$$q_{\text{disk-top}} \approx 2(\theta_{co} - \theta_t) a^2 \varepsilon_a \sigma_B T_d(r)^4 , \quad (B11)$$

where $\theta_{co}$ and $\theta_t$ are defined in Figure B2.

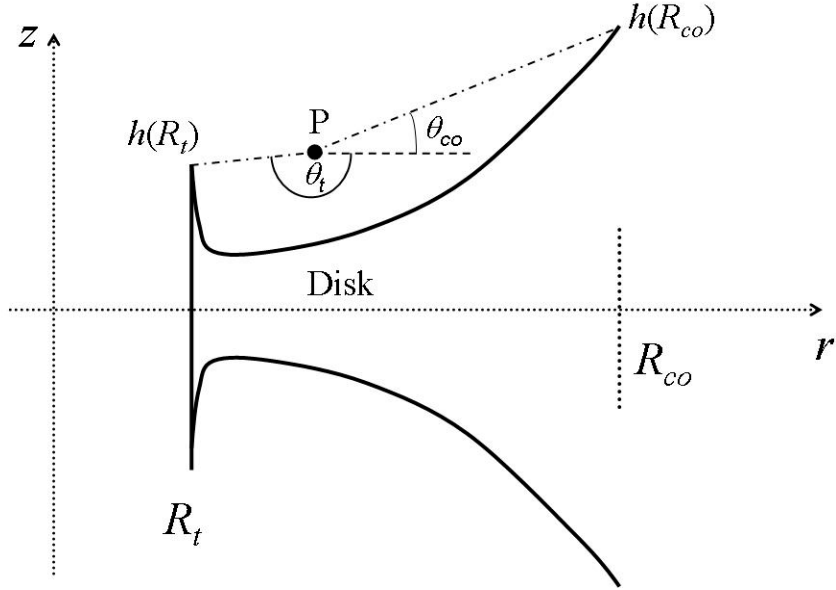

Figure B2: Geometry for disk radiation. A particle is located at P and is receiving radiation from the disk between the truncation radius and the co-rotation radius. $\theta_t$ is the angle between the zero line and the highest point of the inner accretion disk: $h(R_t)$, while $\theta_{co}$ is the angle between the zero line and the height of the disk at the co-rotation radius $h(R_{co})$.

**Case 2: $r < R_t$**

Suppose a particle of radius $a$ is located at or near the inner wall of a disk surrounding a star. The particle will, on one side, be exposed to the direct radiation from the star and, on the other side, the diffuse radiation from the disk. The rate at which energy enters the particle from direct radiation is, from Eqn (B4),

$$q_{\text{star-direct}} \approx \frac{(L_* + L_a) a^2 \varepsilon_a}{4R^2} , \quad (B12)$$

We model the diffuse radiation from the front of the disk as a "wall of radiation" (Vinković et al., 2006), where, for an optically thick disk, radiation coming from and going into the disk rim is in equilibrium and the disk rim provides the same radiation flux as the star. For such a model, we can write:

$$q_{rim} \approx \frac{(L_* + L_a)\theta_w a^2 \varepsilon_a}{2\pi R^2}, \tag{B13}$$

where $\theta_w$ is the angle subtended by the wall, as seen by the particle. From the geometry of the "radiation wall" concept, it is possible to show that (Figure B3)

$$\theta_w \approx \cos^{-1}\left(\frac{z^2 + \zeta^2 - h^2}{\sqrt{(z^2 - h^2)^2 + 2\zeta^2(z^2 + h^2) + \zeta^4}}\right), \tag{B14}$$

where, if we assume that the radiation wall is near or at the inner rim of the disk then

$$\zeta = R_t - r. \tag{B15}$$

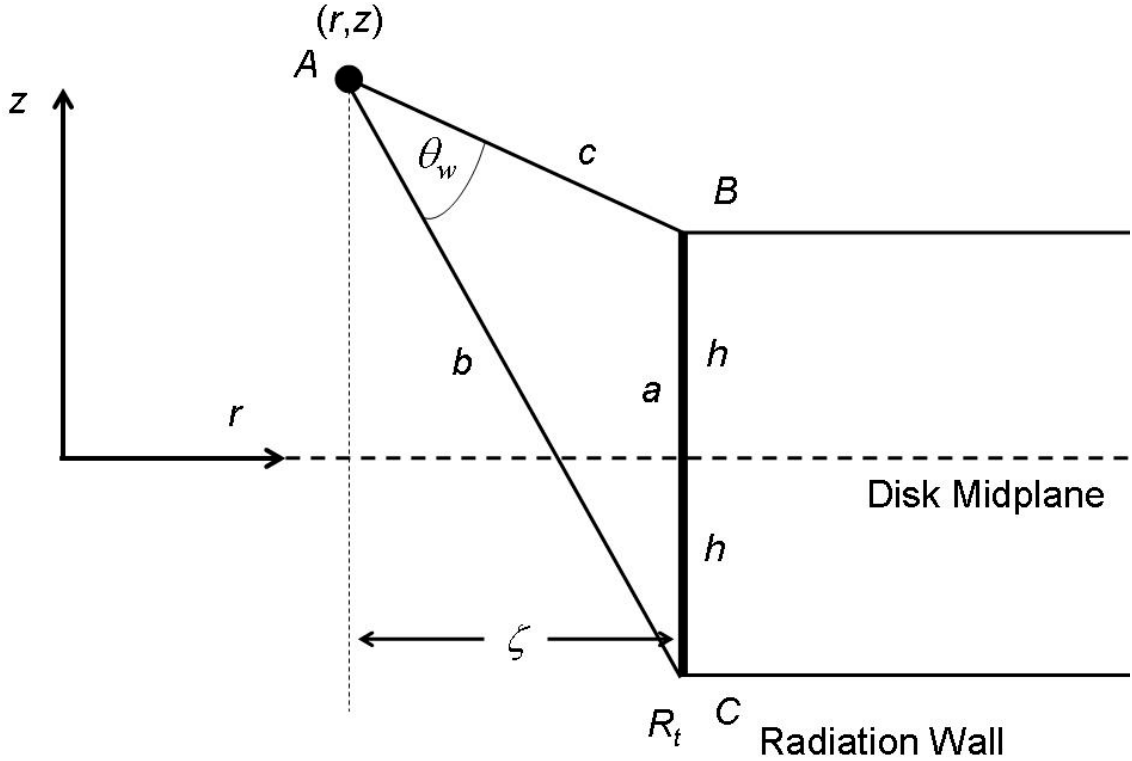

Figure B3: Geometry of particle irradiation from the inner rim of an accretion disk.

Particle Temperature

The energy emitted by the particle, $E_e$, is

$$E_e = 4\pi a^2 \sigma_B \varepsilon_e T_p^4, \tag{B16}$$

where $T_p$ is the temperature of the particle and $\varepsilon_e$ is the particle radiation emission coefficient. If the particle is in radiative equilibrium then the energy absorbed by the particle is equal to the energy emitted by the particle

$$E_e = q_{total} = q_{star} + q_{rim} + \ldots, \tag{B17}$$

so

$$T_p = \left( \frac{q_{total}}{4\pi a^2 \varepsilon_e \sigma_B} \right)^{1/4}, \tag{B18}$$

For the special case of direct, unobscured radiation from the star, we can combine Eqns (B12) and (B17) to give

$$T_{p-star-direct} = \left( \frac{(L_* + L_a)\varepsilon_a}{16\pi R^2 \sigma_B \varepsilon_e} \right)^{1/4}, \tag{B19}$$

Radiation Pressure Direction -Stellar Radiation

For a particle that is subject to radiation pressure from the full disk of a star, it is relatively straightforward to deduce the direction of the radiation pressure. The situation becomes more complex, however, when the stellar disk is partially obscured. In Figure B4, we show the geometry for radiation received by a particle, $P$, from a point on the surface of the star. For the unit vector, $\hat{l}$, between the particle and the star, where

$$\hat{l} = (-\cos\theta, \sin\theta\sin\phi, \sin\theta\cos\phi), \tag{B20}$$

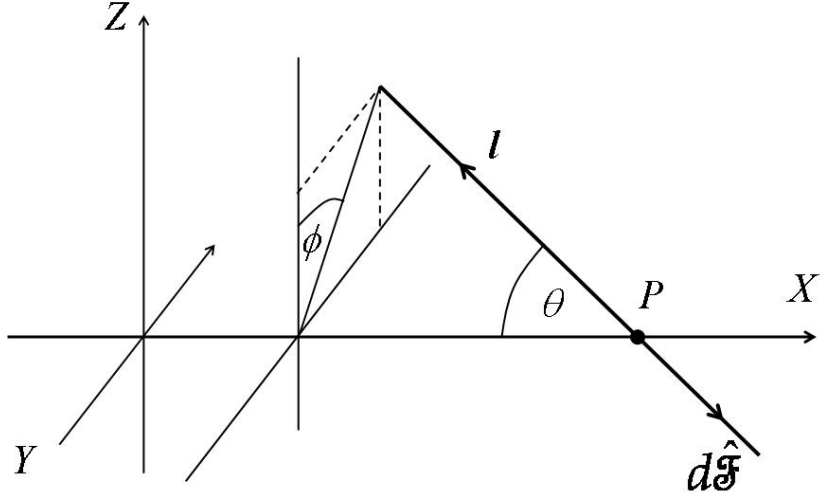

Figure B4: Geometry of a differential element of stellar radiation observed from a particle, $P$, which allows the computation of the radiation pressure direction, $d\hat{\mathscr{F}}$.

The flux of radiation from an element of area is the intensity, $I$, times the solid angle from the point $P$. So, the radiation flux vector from the upper half of the solar disk, $\mathscr{F}_u$, is

$$\mathcal{F}_u \approx \int_{-\pi/2}^{\pi/2} I \int_{\theta_{min}}^{\theta_{max}} \left(\cos\theta, -\sin\theta\sin\phi, -\sin\theta\cos\phi\right)\sin\theta \, d\theta d\phi$$
$$= I\left(\frac{\pi}{4}\left(\cos 2\theta_{min} - \cos 2\theta_{max}\right), 0, \theta_{min} - \theta_{max} + \frac{1}{2}\left(\sin 2\theta_{max} - \sin 2\theta_{min}\right)\right) \quad \text{(B21)}$$

where $I$ is assumed to be a constant $\theta_{min} \geq 0$ and

$$\theta_{max} = \sin^{-1}\left(\frac{R_\bullet}{R}\right). \quad \text{(B22)}$$

Similarly, for the lower part of the solar disk, the geometry is simply reflected through the $XY$ plane of Figure B4 and the radiation flux vector from the lower half of the solar disk, $\mathcal{F}_l$, is

$$\mathcal{F}_l \approx \int_{-\pi/2}^{\pi/2} I \int_0^{\theta_m} \left(\cos\theta, \sin\theta\sin\phi, \sin\theta\cos\phi\right)\sin\theta \, d\theta d\phi$$
$$= I\left(\frac{\pi}{2}\sin^2\theta_m, 0, \theta_m - \frac{\sin 2\theta_m}{2}\right) \quad \text{(B23)}$$

where $0 \leq \theta \leq \theta_m \leq \theta_{max}$. A star that is still visible below the centre line has a total radiation flux vector, $\mathcal{F}_T$, given by

$$\mathcal{F}_T = \mathcal{F}_u\big|_{\theta_{min}=0} + \mathcal{F}_l \approx$$
$$I\left(\frac{\pi}{2}\left(\sin^2\theta_{max} + \sin^2\theta_m\right), 0, \theta_m - \theta_{max} + \frac{\sin 2\theta_{max} - \sin 2\theta_m}{2}\right). \quad \text{(B24)}$$

Returning to Figure B1, we see that an observer/particle located at the point with polar coordinates $(r, z)$ will see the centre-line of the star rotated at an angle $\phi$ to the midplane of the disk. From Eqns (B21), (B23) and (B24), it is apparent that the $Y$ component of the radiation flux has a value of zero and we can convert these equations to polar co-ordinates with the $x$ value corresponding to the $r$ value. Hence the radiation flux directions for a particle at an angle $\phi$ to the midplane of the disk are

$$\mathcal{F}'_{Tr'} = \mathcal{F}_{Tr}\cos\phi - \mathcal{F}_{Tz}\sin\phi = \frac{r\mathcal{F}_{Tr}}{R} - \frac{z\mathcal{F}_{Tz}}{R}$$
$$\mathcal{F}'_{Tz'} = \mathcal{F}_{Tr}\sin\phi + \mathcal{F}_{Tz}\cos\phi = \frac{z\mathcal{F}_{Tr}}{R} + \frac{r\mathcal{F}_{Tz}}{R}, \quad \text{(B25)}$$

where

$$\mathcal{F}_{Tr} = \frac{\pi}{2}I\left(\sin^2\theta_{max} + \sin^2\theta_m\right)$$
$$\mathcal{F}_{Tz} = \theta_m - \theta_{max} + \frac{\sin 2\theta_{max} - \sin 2\theta_m}{2}. \quad \text{(B26)}$$

Disk Radiation

In Figure B2, a particle located at the point P has a polar co-ordinate unit vector, $\hat{\mathbf{d}}$, originating from P and pointing towards an element of the disk. That disk element will produce a radiation pressure force in the opposite direction, $\hat{\mathbf{p}}$, where

$$\hat{\mathbf{p}} = -\hat{\mathbf{d}} = -\cos\theta\,\mathbf{i} - \sin\theta\,\mathbf{j}, \tag{B27}$$

here, **i** and **j** are the unit vectors in the *r* and *z* directions, respectively. The total radiation pressure direction from the disk is

$$\begin{aligned}\mathbf{p}_{total} &\approx -\int_{\theta_t}^{\theta_{co}} \cos\theta\,\mathbf{i} + \sin\theta\,\mathbf{j}\,d\theta \\ &= (\sin\theta_t - \sin\theta_{co})\mathbf{i} + (\cos\theta_{co} - \cos\theta_t)\mathbf{j}\end{aligned}. \tag{B28}$$

**Appendix C**

**Toroidal Magnetic Field Above and Below the Inner Solar Accretion Disk**

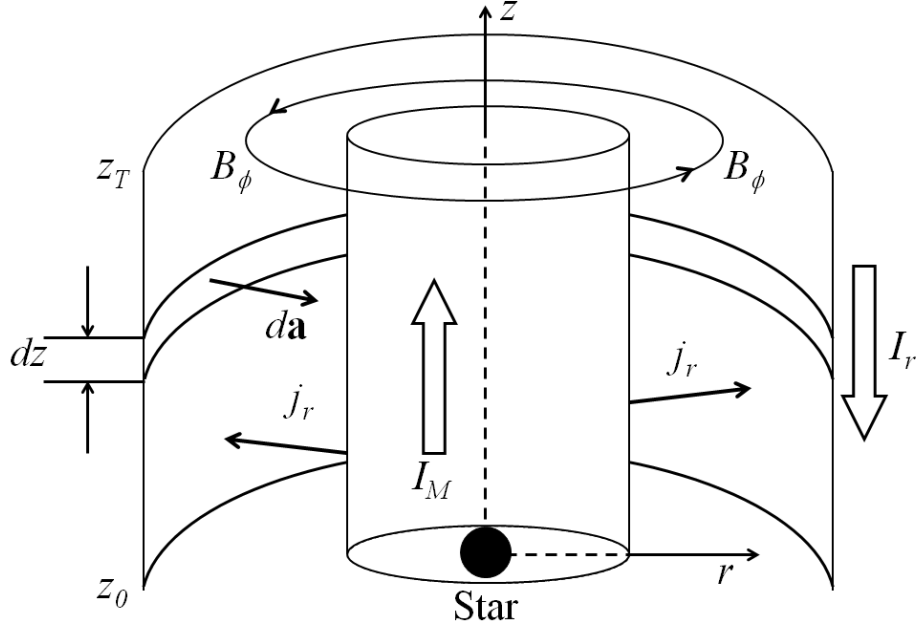

Figure C1: Geometry of the current flow in the jet acceleration region. $I_M$ is the current between the inner edge of the disk and the star; $j_r$ is the radial transfield current density between the stellar field lines; $I_r$ is the total radial transfield current at a distance $r$ from the star; $B_\phi$ is the tordoidal magnetic field generated by $I_M$. This picture is not to scale.

A schematic depiction of the current flow of the toroidal magnetic field is given in Figure . The toroidal field (*i.e.*, a magnetic field that lies in a plane parallel to the plane of the accretion disk) region occurs between an altitude $z_0$ and $z_T$. In this case, a transfield radial current density, $j_r$, bleeds off the magnetospheric current, $I_M$, between the star and the inner disk (it should be noted that $I_M$ may have suffered prior current loss due to current flowing into the star). The toroidal magnetic field within this region is given by Ampere's Law:

$$\nabla \times \mathbf{B} = \mu_0 \mathbf{j} . \tag{C1}$$

Integrating the right hand side of Eqn (C1) over the area element *d*a shown in Figure gives

$$\int_A \mu_0 \mathbf{j} \cdot d\mathbf{a} = -2\pi\mu_0 r \int_{z_0}^{z} j_r(z) dz = -2\mu_0 \pi r j_r(z_c)(z - z_0) , \tag{C2}$$

where the final equality is derived from the Mean Value Theorem with $z_c \in [z_0, z]$. The average radial current $\overline{j_r}$ is given by the definition

$$\overline{j_r} = \frac{1}{z_T - z_0} \int_{z_0}^{z_T} j_r(z) dz = \frac{I_r}{2\pi r (z_T - z_0)} , \tag{C3}$$

where $I_r$ is the total amount of radial current leaking from the magnetospheric current flow $I_M$, so we can write

$$\int_A \mu_0 \, \mathbf{j} \cdot d\mathbf{a} = -\mu_0 I_r \frac{j_r(z_c)}{\bar{j}_r} \frac{(z-z_0)}{(z_T-z_0)}. \tag{C4}$$

Integrating the left hand side of Eqn (C1) over the boundary of the area element $d\mathbf{a}$ shown in Figure gives

$$\int_A \nabla \times \mathbf{B} \cdot d\mathbf{a} = \oint_{\partial A} \mathbf{B} \cdot d\mathbf{l} = 2\pi r (B_\phi(z) - B_\phi(z_0)), \tag{C5}$$

where we note that

$$B_\phi(z_0) = \frac{\mu_0 I_M}{2\pi r}. \tag{C6}$$

Combining these equations gives the general result (assuming axisymmetry)

$$B_\phi(z) = \frac{\mu_0 I_M}{2\pi r}\left(1 - \left(\frac{I_r}{I_M}\right)\left(\frac{j_r(z_c)}{\bar{j}_r}\right)\left(\frac{z-z_0}{z_T-z_0}\right)\right). \tag{C7}$$

It follows that

$$B_\phi(z_T) = \frac{\mu_0 I_M}{2\pi r}\left(1 - \left(\frac{I_r}{I_M}\right)\right), \tag{C8}$$

since $j_r(z_c) = \bar{j}_r$ when $z_c \in [z_0, z_T]$. If $I_r = I_M$ then $B_\phi(z_T) = 0$.

Finally, if we make the approximations that $I_r \approx I_M$ and $j_r(z_c) \approx \bar{j}_r \; \forall z_c \in [z_0, z]$ then Eqn (C7) has the simplified form

$$B_\phi(z) \approx \frac{\mu_0 I_M}{2\pi r}\left(1 - \left(\frac{z-z_0}{z_T-z_0}\right)\right). \tag{C9}$$

**Appendix D**

## $I_M$ Disk-Magnetosphere Half Current at $R_t$

The interaction between the stellar magnetosphere with the accretion disk produces an electric field within the disk (Figure 7). In this case, the current flows from the disk, $\mathbf{j}_D$, to the star via the magnetospheric current density, $\mathbf{j}_M$, and returns along the outer stellar field lines, $j_z$. The magnetospheric current, $\mathbf{I}_M$, is the total current generated in the top or bottom half of the disk that travels between the disk and the star.

We can deduce $I_M$ by using the disk current density, $\mathbf{j}_D$, as given in Liffman (2007):

$$j_D(r) = \sigma_D(r) r \Omega_K(r) B_{*z}(r) \left[1 - \left(\frac{r}{R_{co}}\right)^{3/2}\right], \tag{D1}$$

where $\sigma_D(r)$, the conductivity of the inner disk (Heyvaerts et al., 1996):

$$\sigma_D \approx \frac{1}{\mu_0 \nu}, \tag{D2}$$

with the kinematic viscosity, $\nu$, of the inner disk satisfying

$$\nu \approx \alpha c_s h, \tag{D3}$$

where $\alpha$ is a non-dimensional parameter with a value $\lesssim 1$ and $c_s$ is the sound speed given by

$$c_s = \sqrt{\frac{\gamma p}{\rho}} = \sqrt{\frac{\gamma k_B T}{\bar{m}}}, \tag{D4}$$

$\gamma$ - the ratio of the specific heats. Using Eqns (D1), (D2), (D3) and (22), we can determine $I_M$:

$$I_M = 2\pi R_t h |j_D(R_t)| \approx \frac{2\pi R_t B_{*z}(R_t)}{\alpha \mu_0} \left(\frac{v_K(R_t)}{c_s}\right) \left[1 - \left(\frac{R_t}{R_{co}}\right)^{3/2}\right], \tag{D5}$$

where $v_K$ is the Keplerian speed of a particle around a star:

$$v_K(r) = \sqrt{\frac{GM_*}{r}}. \tag{D6}$$

**Appendix E**

**RMN Formation Timescales**

It is useful to consider the formation timescales of RMNs. A RMN of radius $a_p$ and density $\rho_p$ condensing out of a (Maxwell-Boltzmann) gas will grow in size when atoms from the gas collide and stick to the particle. It is possible to show that

$$\frac{da_p}{dt} = \frac{\sum_{i=1}^{N} Q_i(t) n_{gi}(t) m_i \bar{v}_{gi}(t)}{4\rho_p(t)}, \tag{E1}$$

here, "$i$" denotes the $i$th species of a total of $N$ types of atoms or molecules that are colliding with the condensing particle, $Q_i$ is the "sticking" coefficient ($Q_i = 0$, no sticking, $Q_i = 1$, all colliding gas $i$-type atoms or molecules stick to the condensing particle), $n_{gi}$ is the number density of the $i$th species in the gas phase, $m_i$ is the mass of the $i$-type atom or molecule and $\bar{v}_{gi}\left(=\sqrt{8k_B T_g / \pi m_i}\right)$ the average velocity of the gaseous $i$-type atom or molecule at a temperature $T_g$. Using the pressure $p_i$ of the $i$-th gas component, the ideal gas law gives

$$\frac{da_p}{dt} = \frac{1}{\rho_p(t)\sqrt{2\pi k_B T_g(t)}} \sum_{i=1}^{N} Q_i(t) p_i(t) \sqrt{m_i}. \tag{E2}$$

Due to the time dependent nature of the quantities in Eqn (E2), it is difficult to obtain a timescale for condensation. It is a little more tractable if we consider one of the condensing species. In this case, Eqn (E1), becomes:

$$\frac{da_p}{dt} = \frac{Q_i(t) n_{gi}(t) m_i \bar{v}_{gi}(t)}{4 f_i(t) \rho_p(t)}, \tag{E3}$$

with $f_i(t)$ the mass fraction of the condensing particle made of the $i$th substance. From this equation, it is possible to define the timescale for doubling the radius of the particle from an initial value of $a_{p0}$ to $2a_{p0}$:

$$\tau_{pdi} \approx \frac{f_i \rho_p a_{p0}}{Q_i X_i p_T} \sqrt{\frac{2\pi k_B T_g}{m_i}}$$

$$\approx 68 \frac{\left(\dfrac{f_i}{0.2274}\right)\left(\dfrac{\rho_p}{17.4\,\text{g cm}^{-3}}\right)\left(\dfrac{a_{p0}}{200\,\text{nm}}\right)}{\left(\dfrac{Q_i}{1}\right)\left(\dfrac{X_i}{2.4\times 10^{-11}}\right)\left(\dfrac{p_T}{10\,\text{Pa}}\right)} \sqrt{\dfrac{\left(\dfrac{T_g}{1600\,\text{K}}\right)}{\left(\dfrac{m_i}{190.3\,\text{amu}}\right)}} \;\text{yrs}. \tag{E4}$$

Here, $X_i = \dfrac{n_i}{n_T}$ with $n_T$ being the number density of the gas and $p_T$ the total pressure of the gas. The nominal value of 0.2274 for $f_i$ was obtained from Berg et al. (2009) for the average mass fraction of Os in RMNs, while the value of 17.4 g cm$^{-3}$ for $\rho_p$ has been deduced from Berg et al. (2009) by using the average mass fractions for the elements that constitute RMNs: Ir, Mo, Fe, Pt, W, Os, Ru, Rh and Ni. The value of $2.4\times 10^{-11}$ for $X_i$ is the solar abundance for Os (Kerridge & Matthews, 1988). The deduced time scale of RMN formation is of order one hundred years, which is consistent with the timescales obtained from Berg et al. (2009).

The formation timescales increase with decreasing pressure. The pressures obtained in the accretion channel (Figure 6) for ages of the Solar Nebula in excess of 400,000 years give RMN formation timescales in the range of $10^4$ to $10^6$ years. Such improbably long formation timescales are consistent with the idea that RMN formed in the first 100,000 years of the Solar Nebula (§3), when the pressures in the accretion columns were near their maximum values.

It should be noted that the calculation given here is an order-of-magnitude estimate. A more comprehensive and accurate calculation is given in (Petaev et al., 2003), who also modify the sticking coefficients so that mass balance is maintained for the different elements between the condensed and gaseous phases.

**Appendix F**

**RMN Charge**

A RMN or dust grain in the inner Solar Nebula may become charged due to collisions with electrons, protons and other charged particles. Radiation exposure from the proto-Sun and inner disk also allows the possibility of charging the grain via the photo-ejection of electrons.

As discussed in Bellan (2006) and Tielens (2005), a number of different current flows can produce charged dust grain. One source of current is the "attractive" flow of charged particles onto the dust grain, where they have an opposite charge to the dust grain charge:

$$I_{att} = \frac{2}{\sqrt{\pi}} \left(1 - \frac{q\phi_d}{k_B T_q}\right) q n_{0q} \pi r_d^2 v_{Tq} \,, \tag{F1}$$

with $q$ the charge on the particle colliding with the dust grain, $T_q$ is the temperature of the gas of charged particles with charge $q$, $\phi_d$ is the electric potential of the dust grain:

$$\phi_d \approx \frac{-Z_d e}{4\pi\varepsilon_0 r_d} \,, \tag{F2}$$

$e$ is the electric charge, $Z_d$ is the number of charges on the dust grain, $n_{0q}$ the gas number density of the "$q$" charges colliding with the dust particle, $r_d$ the radius of the dust grain, $\varepsilon_0$ the permittivity of free space and $v_{Tq}$ is the thermal speed of the $q$ charged particles. The thermal speed for an "$i$"-th type of gas particle is given by

$$v_{Ti} = \sqrt{\frac{2k_B T_i}{m_i}} \,, \tag{F3}$$

with $m$ the mass of the gas particle.

The "repulsive" current flow for the gas particles with the same charge sign as the dust particle has the form:

$$I_{rep} = \frac{2}{\sqrt{\pi}} \eta n_{0\eta} v_{T\eta} \pi r_d^2 \exp\left(\frac{-\eta \phi_d}{k_B T_\eta}\right) , \tag{F4}$$

where, in this case, we have assumed that the charged particle has a charge "$\eta$".

In many cases of interest, the dust particle will have a negative charge due to the much higher thermal speed of the electron. In this paper, we assume a pure hydrogen atmosphere. As a consequence, the positive and negative charges will be, respectively, protons and electrons with $q = e$ and $\eta = -e$. We also assume that the number density of dust particles, $n_d$, is small relative to the number densities of electrons and protons:

$$n_{ion} \approx n_e \,, \tag{F5}$$

where we note that for a significant number density of dust particles, Eqn (F5) becomes

$$n_{ion} = n_e + Z_d n_d \,. \tag{F6}$$

In such a situation, the dust particles may come in close proximity and form crystalline-like structures due to their mutually repulsive charges (Fortov, 2004). The repulsive and attractive particles will reach their equilibrium flow rates when $I_{att} = I_{rep}$, which implies

$$\exp\left(\frac{e\phi_d}{k_B T}\right) \approx \sqrt{\frac{m_e}{m_p}}\left(1 - \frac{e\phi_d}{k_B T}\right), \tag{F7}$$

and where we have assumed that the proton temperature is approximately equal to the electron temperature. It can be shown numerically that Eqn (F7) has the semi-analytic solution:

$$Z_d \approx \frac{-2.504 \times 4\pi\varepsilon_0 r_d k_B T}{e^2} \approx -67.5\left(\frac{r_d}{0.3\,\text{micron}}\right)\left(\frac{T}{1500\,K}\right). \tag{F8}$$

An additional current term arises when the dust particle is subject to radiation with photon energies high enough to photo-eject electrons from the dust grain. The photo-ejection current to the dust grain is

$$I_{pe} = 4\pi e \int_{v_{Zd}}^{\infty} \frac{J(v)}{hv}\pi r_d^2 Y_{pe}(Z_d, v)\, dv, \tag{F9}$$

with $J(v)$ the mean intensity of radiation, $h$ is Planck's constant, $v$ the radiation frequency and $Y_{pe}$ the photo-electric yield. The term $4\pi J(v)$ is the radiative flux, i.e, the amount of radiative energy that is passing through a unit area per unit time per unit frequency interval. For a particle in free space near a star of radius $R_*$

$$4\pi J(v) = \frac{L(v)}{4\pi R^2} \approx \pi\left(\frac{R_*}{R}\right)^2 B_v(T_{eff}) = \pi\left(\frac{R_*}{R}\right)^2 \frac{2hv^3}{c^2\left(\exp\left(\frac{hv}{k_B T_{eff}}\right) - 1\right)}, \tag{F10}$$

with $L(v)$ the luminosity of the radiation source as a function of frequency, $B_v(T)$ is the Planck blackbody function and $T_{eff}$ is the effective temperature:

$$T_{eff} = \left(\frac{L}{4\pi R_*^2 \sigma_B}\right)^{\frac{1}{4}}, \tag{F11}$$

with $L$ the total luminosity of the source (*i.e.* stellar and accretion luminosities). The photo-electric yield is quite a complex quantity and is dependent on the size and composition of the dust grains (e.g., Weingartner and Draine, 2001). We have no direct information regarding the photo-electric yield of RMNs, so, as an initial approximation, we will adopt the formulation of Tielens (2005):

$$Y_{pe}(Z_d, v) = Y_\infty\left(1 - \frac{W(Z_d)}{hv}\right)f(r_d), \tag{F12}$$

with $Y_\infty$ the photoelectric yield for bulk materials, $W(Z_d)$ the work function (which typically has a value of 4 to 6 eV) and $f(r_d)$ is the yield enhancement factor for small particles. Following Tielens (2005), we set $f(r_d) = 1$, $Y_\infty = 0.15$ and $W(Z_d) = 5$ eV.

The lower limit on the integral in Eqn (F9) is the photon frequency required to eject electrons from the RMN:

$$\nu_{Z_d} = \begin{cases} \nu_c + \dfrac{Z_d e^2}{4\pi\varepsilon_0 h r_d} & Z_d > 0 \\ \nu_c & Z_d < 0 \end{cases}, \quad (F13)$$

where

$$\nu_c = \frac{W(Z_d)}{h}. \quad (F14)$$

So the photoejection current for an RMN with an unobstructed view of the proto-Sun is

$$I_{pe} \approx \frac{0.3 e \pi^2 r_d^2 R_*^2}{c^2 R^2} \int_{\nu_{Zd}}^{\infty} \frac{\nu^2 (1 - 5\,\text{eV}/h\nu)}{(\exp(h\nu/k_B T_{eff}) - 1)} d\nu . \quad (F15)$$

For the case $\phi \geq \theta_h$ (see Appendix B)

$$I_{pe} \approx \frac{0.15 e \pi^2 r_d^2 R_*^2}{c^2 R^2} \int_{\nu_{Zd}}^{\infty} \frac{\nu^2 (1 - 5\,\text{eV}/h\nu)}{(\exp(h\nu/k_B T_{eff}) - 1)} \left(1 + \left(\frac{R}{R_*}\right)^2 \sin^2 \psi_l \right) d\nu . \quad (F16)$$

Similarly, for the case $\phi < \theta_h$

$$I_{pe} \approx \frac{0.15 e \pi^2 r_d^2 R_*^2}{c^2 R^2} \int_{\nu_{Zd}}^{\infty} \frac{\nu^2 (1 - 5\,\text{eV}/h\nu)}{(\exp(h\nu/k_B T_{eff}) - 1)} \left(1 - \left(\frac{R}{R_*}\right)^2 \sin^2 \psi_l \right) d\nu . \quad (F17)$$

To find $Z_d$, we numerically solve the equation

$$I_{att} + I_{rep} + I_{pe} = 0 . \quad (F18)$$